\documentclass[onecolumn]{IEEEtran}
\usepackage{amsthm, amsmath, amssymb, amsfonts, url, booktabs, tikz, setspace, fancyhdr, bm, bbm}
\usepackage{multirow}
\usepackage{booktabs}
\usepackage{algorithmic}
\usepackage[all,pdf]{xy}
\usepackage[ruled,linesnumbered]{algorithm2e}
\usepackage{array}
\usepackage{arydshln}
\usepackage[justification=centering,caption=false,font=normalsize,labelfont=sf,textfont=sf]{subfig}
\usepackage{textcomp}
\usepackage{stfloats}
\usepackage{comment}
\usepackage{verbatim}
\usepackage{graphicx}
\usepackage[noadjust]{cite}
\usepackage{xcolor}
\usetikzlibrary{matrix}
\usepackage{hyperref}
\newtheorem{theorem}{Theorem}[section]
\newtheorem{definition}[theorem]{Definition}
\newtheorem{lemma}[theorem]{Lemma}

\newtheorem{remark}[theorem]{Remark}
\newtheorem{observation}[theorem]{Observation}

\begin{document}

\title{Correcting Errors Through Partitioning and Burst-Deletion Correction}

\author{Yubo~Sun and Gennian~Ge
\thanks{This research was supported by the National Key Research and Development Program of China under Grant 2020YFA0712100, the National Natural Science Foundation of China under Grant 12231014, and Beijing Scholars Program.}
\thanks{Y. Sun ({\tt 2200502135@cnu.edu.cn}) and G. Ge ({\tt gnge@zju.edu.cn}) are with the School of Mathematical Sciences, Capital Normal University, Beijing 100048, China.}
}

\maketitle

\begin{abstract}
In this paper, we propose a partitioning technique that decomposes a pair of sequences with overlapping $t$-deletion $s$-substitution balls into sub-pairs, where the $^{\leq}t$-burst-deletion balls of each sub-pair intersect. 
This decomposition facilitates the development of $t$-deletion $s$-substitution correcting codes that leverage approaches from $^{\leq}t$-burst-deletion correction. 
Building upon established approaches in the $^{\leq}t$-burst-deletion correction domain, we construct $t$-deletion $s$-substitution correcting codes for $t\in \{1,2\}$ over binary alphabets and for $t=1$ in non-binary alphabets, with some constructions matching existing results and others outperforming current methods.
Our framework offers new insights into the underlying principles of prior works, elucidates the limitations of current approaches, and provides a unified perspective on error correction strategies.
\end{abstract}

\section{Introduction}

\IEEEPARstart{D}{ELETIONS}, insertions, and substitutions represent the most prevalent error types affecting data stored in DNA-based systems \cite{Organick-18-background, Yazdi-15-TMBMC-DNA}. 
The development of explicit codes to address these errors is both a natural and critical endeavor. 
It is well-known in \cite{Smagloy-23-IT-DS} that the correction capability for $t$ deletions and $s$ substitutions is equivalent to that of handling a combination of $t$ deletions and insertions,  alongside $s$ substitutions. 
Consequently, it suffices to concentrate efforts on correcting deletions and substitutions.

One of the primary challenges in error correction lies in addressing deletions, which often disrupt synchronization between the sender and receiver.
This challenge becomes even more pronounced when deletions are coupled with other error types. 
According to existing literature, there are essentially three main approaches for constructing low-redundancy codes capable of correcting deletion-related errors. 
The first approach employs Varshamov-Tenengolts (VT) syndromes \cite{Guruswami-21-IT-2D, Liu-24-IT, Song-23-IT-BD, Pi-24-arXiv-2E, Sima-21-IT-kD, Sun-24-IT-2E, Sima-20-ISIT-tD, Song-22-IT-DS, Sima-20-IT-2D, Ye-24-arXiv-BD}. 
This method is particularly attractive due to its support for linear-time encoding, making it highly efficient for practical applications. 
The second approach introduces a series of markers to transform the correction of deletions into the problem of correcting errors within the Hamming metric \cite{Brakensiek-18-IT-kD, Gabrys-19-IT-2D}. However, the redundancy of this method depends on the number of marker types and may be higher than that of the first approach.
The third approach leverages syndrome compression techniques or Linial's coloring algorithm (with pre-coding) \cite{Li-23-DS, Sima-20-ISIT-q, Sima-20-ISIT, Sima-20-ISIT-tD, Song-22-IT-DS}. 
While this strategy has the potential to achieve lower redundancy compared to the first method, it is less likely to maintain the desirable linear complexity in its encoding algorithms.
In this paper, we focus on the first approach, utilizing VT syndromes to develop efficient error-correcting codes with linear-time encoding complexity.
Our aim is to elucidate the core principles underlying prior research, clarify the limitations of existing methods, and propose a comprehensive framework for error correction strategies.

Two primary approaches for constructing error-correcting codes based on VT syndromes involve utilizing either the original sequences or their corresponding differential sequences. 
A notable result in this area is attributed to Levenshtein \cite{Levenshtein-66-SPD-1D}, who successfully utilized VT syndromes applied directly to the original sequences to construct binary codes capable of correcting a single deletion. 
Building upon this foundation, subsequent research by Smagloy et al. \cite{Smagloy-23-IT-DS} and Song et al. \cite{Song-22-IT-DS} extended this method to address binary codes correcting one deletion and $s$ substitutions, for $s=1$ and $s \geq 1$, respectively.
However, this approach has significant limitations. Specifically, it encounters challenges in correcting multiple deletions in binary alphabets \cite{Sima-20-IT-2D, Sima-21-IT-kD} and even a single deletion correction in non-binary alphabets \cite{Tenengolts-84-IT-q_D, Gabrys-23-IT-DS, Cai-21-E, Sun-25-IT, Nguyen-24-IT-1D}.
Shifting focus to the application of VT syndromes to differential sequences, Levenshtein \cite{Levenshtein-70-BD} derived binary codes capable of correcting a burst of at most two deletions. Further advancements were made by Nguyen et al. \cite{Nguyen-24-IT-1D}, who designed non-binary codes for correcting a single deletion, and Pi and Zhang \cite{Pi-24-arXiv-2E}, who developed binary codes that can correct a combination of at most two deletions and substitutions. \footnote{Pi and Zhang \cite{Pi-24-arXiv-2E} constructed codes by applying VT syndromes to run strings, introduced in \cite{Levenshtein-70-BD, Guruswami-21-IT-2D}, which is equivalent to applying VT syndromes to differential sequences \cite{Sun-24-IT-2E}.} 
It remains an open problem whether the application of VT syndromes to differential sequences can be extended to correct three deletions in binary alphabets or two deletions in non-binary alphabets.
In addition to these advancements, recent research \cite{Guruswami-21-IT-2D, Sun-24-IT-2E} has explored the application of VT syndromes to multiple sequence types, which is beyond the scope of this paper.

An intriguing phenomenon is that, in binary alphabets, the existing methods designed to correct a burst of at most two deletions can be extended to correct any arbitrary two deletions. 
Meanwhile, no efficient solutions are currently available to correct a burst of at most three deletions in binary alphabets or a burst of at most two deletions in non-binary alphabets, and effective approaches for correcting these as arbitrary deletions remain elusive in the literature. 
Intuitively, one might expect that the problem of correcting a burst of at most $t$ deletions is closely related to that of correcting $t$ arbitrary deletions. 

In this paper, we introduce a novel framework that establishes a definitive link between error correction and burst-deletion correction, thereby validating this intuition. 
The key insight lies in the decomposition of a pair of sequences with overlapping $t$-deletion $s$-substitution balls into sub-pairs whose $^{\leq}t$-burst-deletion balls intersect.
Consequently, if certain classes of sequences can be effectively employed to correct a burst of at most $t$ deletions with desirable properties, they may also be applicable to correcting $t$ arbitrary deletions along with any number of substitutions.
Within this framework, we provide concise proofs of existing results and derive error-correcting codes capable of managing a wider range of errors. 
Specifically, by employing VT syndromes to the original sequences (in their accumulative form), we construct binary single-deletion $s$-substitution correcting codes with $(s+1)(2s+1)\log n + O(1)$ bits of redundancy, matching the results of Levenshtein \cite{Levenshtein-66-SPD-1D} for $s=0$, Smagloy et al. \cite{Smagloy-23-IT-DS} for $s=1$, and Song et al. \cite{Song-22-IT-DS} for $s \geq 1$, respectively.
By applying VT syndromes to the differential sequences (also in their accumulative form), we construct $q$-ary single-deletion $s$-substitution correcting codes with $(s+1)(2s+1)\log n+O(1)$ bits of redundancy and binary two-deletion $s$-substitution correcting codes with $(s+2)(2s+3)\log n+O(1)$ bits of redundancy, generalizing the work of Nguyen et al. \cite{Nguyen-24-IT-1D} and Pi and Zhang \cite{Pi-24-arXiv-2E}.
Notably, when $s = 1$, this yields $q$-ary single-deletion single-substitution correcting codes with $6\log n + O(1)$ bits of redundancy, outperforming existing methods based on syndrome compression or Linial's algorithm with pre-coding, which require $7\log n + o(\log n)$ bits of redundancy. 
All our binary codes, without any changes, can be used to correct $t$ deletions in combination with $s$ substitutions and adjacent transpositions, maintaining the same parameters $t$ and $s$.
Additionally, we analyze the limitations of applying VT syndromes directly to original sequences for correcting two deletions in binary alphabets and one deletion in non-binary alphabets, as well as the limitations of simply applying VT syndromes to differential sequences for correcting three deletions in binary alphabets and two deletions in non-binary alphabets.
For multiple deletions, Sima et al. \cite{Sima-20-IT-2D, Sima-20-ISIT-tD, Sima-21-IT-kD} demonstrated that applying VT syndromes to binary sequences with adjacent ones spaced at least $t$ apart can correct $t$ deletions. 
We revisit this method and generalize it to arbitrary alphabets, thereby broadening its scope and practical utility.

The rest of this paper is organized as follows. 
Section \ref{sec:pre} introduces relevant notations and definitions used throughout the paper. 
Section \ref{sec:partition} presents a partitioning method to demonstrate that any pair of sequences with overlapping $t$-deletion $s$-substitution balls can be decomposed into sub-pairs, whose $^{\leq}t$-burst-deletion balls have a non-empty intersection. 
Section \ref{sec:code} delves into the construction of codes for correcting either one or two deletions combined with multiple substitutions, including detailed encoding and decoding procedures, and also discusses the limitations of current approaches.
Section \ref{sec:multiple} extends this exploration to codes capable of correcting multiple deletions and multiple substitutions. 
Finally, Section \ref{sec:conclusion} concludes this paper.

\section{Preliminaries}\label{sec:pre}

Let $[i, j]$ denote the set of integers $\{i, i+1, \ldots, j\}$ if $i \leq j$, and the empty set otherwise. For an integer $q \geq 2$, define $\Sigma_q = [0, q - 1]$ as the $q$-ary alphabet. Let $\mathbb{Z}$ be the set of all integers. 
We write $\Sigma_q^n$ (respectively, $\mathbb{Z}^n$) for the set of all sequences of length $n$ over $\Sigma_q$ (respectively, $\mathbb{Z}$). For any sequence $\boldsymbol{x} \in \Sigma_q^n$, we denote it either as $\boldsymbol{x} = x_1 x_2 \cdots x_n$ or as $\boldsymbol{x} = (x_1, x_2, \ldots, x_n)$, where $x_i$ is the $i$-th element of $\boldsymbol{x}$ for $i \in [1, n]$. If $i \notin [1, n]$, we set $x_i = 0$.
Let $\boldsymbol{y} \in \Sigma_q^m$, we define the \emph{concatenation} of $\boldsymbol{x}$ and $\boldsymbol{y}$ as $\boldsymbol{x} \boldsymbol{y} = x_1 x_2 \cdots x_n y_1 y_2 \cdots y_m$. In particular, $\boldsymbol{x}^k$ denotes the substring obtained by concatenating $\boldsymbol{x}$ with itself $k$ times. 
Furthermore, if there exists a set $\mathcal{I} = \{i_1, i_2, \ldots, i_m\}$ with $1 \leq i_1 < i_2 < \cdots < i_m \leq n$, such that $\boldsymbol{y} = \boldsymbol{x}_{\mathcal{I}} \triangleq x_{i_1} x_{i_2} \cdots x_{i_m}$, then $\boldsymbol{y}$ is called a \emph{subsequence} of $\boldsymbol{x}$.
When $\mathcal{I}$ is an interval, we say that $\boldsymbol{x}_{\mathcal{I}}$ is a \emph{substring} of $\boldsymbol{x}$.  
Let $|\boldsymbol{x}|$ be the \emph{length} of $\boldsymbol{x}$ when $\boldsymbol{x}$ is a sequence, and $|\mathcal{S}|$ be the \emph{size} of $\mathcal{S}$ when $\mathcal{S}$ is a set.

Let $\boldsymbol{x}$ be a $q$-ary sequence of length $n$.
\begin{itemize}
    \item A \emph{deletion} at position $i$ in $\boldsymbol{x}$ deletes the $i$-th symbol, yielding the sequence $x_1 \cdots x_{i-1} x_{i+1} \cdots x_n$.
        Moreover, a \emph{$^{\leq}t$-burst-deletion} (or \emph{a burst of at most $t$ deletions}) at position $i$ deletes the substring $\boldsymbol{x}_{[i, i + t' - 1]}$ for some $t' \in [0,t]$, yielding the sequence $x_1 \cdots x_{i-1} x_{i + t'} \cdots x_n$.

    \item A \emph{substitution} at position $i$ in $\boldsymbol{x}$ replaces $x_i$ with some $x_i' \in \Sigma_q$, yielding the sequence $x_1 \cdots x_{i-1} x_i' x_{i+1} \cdots x_n$.
        We allow for trivial substitutions, where $x_i' = x_i$, meaning no actual change occurs.
\end{itemize}
Let $\mathcal{B}_{t,s}(\bm{x})$ be the \emph{$t$-deletion $s$-substitution ball} of $\bm{x}$, consisting of all sequences that can be obtained from $\bm{x}$ after $t$ deletions and $s$ substitutions.  
Similarly, let $\mathcal{D}_{t}(\boldsymbol{x})$ be the \emph{$^{\leq}t$-burst-deletion ball} of $\boldsymbol{x}$, consisting of all sequences obtainable from $\boldsymbol{x}$ after a $^{\leq}t$-burst-deletion.
A code $\mathcal{C}\subseteq \Sigma_q^n$ is called a \emph{$t$-deletion $s$-substitution correcting code} if it can correct $t$ deletions and $s$ substitutions. 
Equivalently, the $t$-deletion $s$-substitution balls centered at any two distinct codewords are disjoint.
If certain type of errors are not considered by the code, they will be omitted from its name.
For example, a code is called a \emph{$t$-deletion correcting code} if it can correct $t$ deletions. 
To evaluate an error-correcting code $\mathcal{C}\subseteq \Sigma_q^{n}$, we calculate its \emph{redundancy}, defined as $n \log q- \log |\mathcal{C}|$, where $\log$ denotes the base-$2$ logarithm.

\section{Partitioning Technique}\label{sec:partition}   

In this section, we propose a partitioning technique demonstrating that any pair of sequences with overlapping error balls can be partitioned into sub-pairs, where the burst-deletion balls of each sub-pair intersect. 
Building upon this, it suffices to focus on codes capable of correcting a burst of deletions.

\begin{theorem}\label{thm:DS}
  For integers $ t\geq 1 $ and $ s\geq 0 $, let $ \bm{x}, \bm{y}$ be sequences with the same length such that $ \mathcal{B}_{t,s}(\bm{x}) \cap \mathcal{B}_{t,s}(\bm{y}) \neq \emptyset $. Then, there exists a partition
  \begin{gather*}
  \bm{x} = \bm{x}^{(1)} \bm{x}^{(2)} \cdots \bm{x}^{(m)},\\
  \bm{y} = \bm{y}^{(1)} \bm{y}^{(2)} \cdots \bm{y}^{(m)},
  \end{gather*}
  with $ m \leq 2t + 2s - 1 $, such that for each $ i \in [1, m] $, $ |\bm{x}^{(i)}| = |\bm{y}^{(i)}| $ and $ \mathcal{D}_{t}(\bm{x}^{(i)}) \cap \mathcal{D}_{t}(\bm{y}^{(i)}) \neq \emptyset$.
\end{theorem}

Before proceeding to the proof, we present the following two lemmas, which will be utilized later.

\begin{lemma}\label{lem:D}
  For any integer $ t \geq 1 $, let $ \bm{x}, \bm{y}$ be sequences of length $n$ such that $ \mathcal{B}_{t,0}(\bm{x}) \cap \mathcal{B}_{t,0}(\bm{y}) \neq \emptyset $. Then, there exists a partition
  \begin{gather*}
  \bm{x} = \bm{x}^{(1)} \bm{x}^{(2)} \cdots \bm{x}^{(m)},\\
  \bm{y} = \bm{y}^{(1)} \bm{y}^{(2)} \cdots \bm{y}^{(m)},
  \end{gather*}
  with $ m \leq 2t - 1 $, such that for each $ i \in [1, m] $, $|\bm{x}^{(i)}| = |\bm{y}^{(i)}|$ and $\mathcal{D}_{t}(\bm{x}^{(i)}) \cap \mathcal{D}_{t}(\bm{y}^{(i)}) \neq \emptyset.$
\end{lemma}

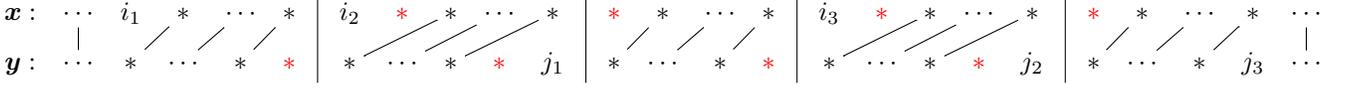
\begin{figure*}
\center
\begin{align*}
\begin{array}{c|c|c|c|c}
\xymatrix@C=0.9ex@R=1.2ex{
\bm{x}: & \cdots \ar@{-}[d] & i_1 & \ast \ar@{-}[dl] & \cdots \ar@{-}[dl] & \ast \ar@{-}[dl] \\
\bm{y}: & \cdots & \ast & \cdots & \ast & {\color{red}\ast}
}
&
\xymatrix@C=0.9ex@R=1ex{
i_2 & {\color{red}\ast} & \ast \ar@{-}[dll] & \cdots \ar@{-}[dll] & \ast \ar@{-}[dll]\\
\ast & \cdots & \ast & {\color{red}\ast} & j_1
}
&
\xymatrix@C=0.9ex@R=1.8ex{
{\color{red}\ast} & \ast \ar@{-}[dl] &\cdots \ar@{-}[dl] & \ast \ar@{-}[dl] \\
\ast &\cdots & \ast & {\color{red}\ast}
}
&
\xymatrix@C=0.9ex@R=1ex{
i_3 & {\color{red}\ast} & \ast \ar@{-}[dll] & \cdots \ar@{-}[dll] & \ast \ar@{-}[dll] \\
\ast & \cdots & \ast & {\color{red}\ast} & j_2
}
&
\xymatrix@C=0.9ex@R=1.4ex{
{\color{red}\ast} & \ast \ar@{-}[dl] & \cdots \ar@{-}[dl]  & \ast \ar@{-}[dl] & \cdots \ar@{-}[d] \\
\ast & \cdots & \ast & j_3 & \cdots
}
\end{array}
\end{align*}
\caption{An illustration example of Lemma \ref{lem:D} when $t=3$. Let $\bm{x}, \bm{y}$ be sequences of length $n$ such that $\bm{x}_{[1,n] \setminus \{i_1,i_2,i_3\}} = \bm{y}_{[1,n] \setminus \{j_1,j_2,j_3\}}$, where $i_1<i_2<j_1<i_3<j_2<j_3$. Each pair of bits connected by a solid line are of equal value.}
\label{fig:deletion}
\end{figure*}

\begin{IEEEproof}
    We prove by induction on $t$.
    For the base case where $t=1$, the conclusion is straightforward.
    Now, assuming that the conclusion is valid for $t<t'$ with $t' \geq 2$, we examine the scenario where $t=t'$.
    Since $\mathcal{B}_{t,0}(\bm{x})\cap \mathcal{B}_{t,0}(\bm{y})\neq \emptyset$, there exist two index sets $\mathcal{I}=\{i_1,\ldots,i_t\}$ and $\mathcal{J}=\{j_1,\ldots,j_t\}$ with $0< i_1<\cdots<i_t< n+1=i_{t+1}$ and $0< j_1<\cdots<j_t< n+1=j_{t+1}$, such that $\bm{x}_{[1,n]\setminus \mathcal{I}}= \bm{y}_{[1,n]\setminus \mathcal{J}}$.
    Let $\ell\in [1,t]$ be the smallest index satisfying  $\max\{i_{\ell},j_{\ell}\}<\min\{i_{\ell+1},j_{\ell+1}\}$.
    We distinguish between two cases.
    \begin{itemize}
      \item If $\ell<t$, let $n'=\max\{i_{\ell},j_{\ell}\}$, it follows that $ \mathcal{B}_{\ell,0}(\bm{x}_{[1,n']}) \cap \mathcal{B}_{\ell,0}(\bm{y}_{[1,n']}) \neq \emptyset $ and $ \mathcal{B}_{t-\ell,0}(\bm{x}_{[n'+1,n]}) \cap \mathcal{B}_{t-\ell,0}(\bm{y}_{[n'+1,n]}) \neq \emptyset $.
      By applying the induction hypothesis on the pairs $(\bm{x}_{[1,n']}, \bm{y}_{[1,n']})$ and $(\bm{x}_{[n'+1,n]}, \bm{y}_{[n'+1,n]})$, respectively, we can conclude that the conclusion holds for this case.
      \item If $\ell=t$, by symmetry, without loss of generality assume $i_1<j_1$, it follows that $i_{k}\geq j_{k+1}$ for $k\in [1,t-1]$.
      We say that $i_k \prec i_{k'}$ and $j_k\prec j_{k'}$ if $i_k< i_{k'}$ and $j_k< j_{k'}$, respectively.
      Moreover, we say that $i_k \prec j_{k'}$ if $i_k\leq j_{k'}$.
      Sort the multi-set $\mathcal{I}\cup \mathcal{J}$ in increasing order: $p_1\prec p_2\prec \cdots \prec p_{2t}$.
      For $k\in [1,2t-1]$ and $\bm{z}\in \{\bm{x},\bm{y}\}$, let
      \begin{equation*}
        \bm{z}^{(k)}=
        \begin{cases}
          \bm{z}_{[p_k,p_{k+1}-1]}, & \mbox{if } p_k,p_{k+1}\in \mathcal{I}; \\
          \bm{z}_{[p_k,p_{k+1}]},   & \mbox{if } p_k\in \mathcal{I}, p_{k+1}\in \mathcal{J};\\
          \bm{z}_{[p_k+1,p_{k+1}-1]},   & \mbox{if } p_k\in \mathcal{J}, p_{k+1}\in \mathcal{I};\\
          \bm{z}_{[p_k+1,p_{k+1}]},   & \mbox{if } p_k, p_{k+1}\in \mathcal{J}.
        \end{cases}
      \end{equation*}
      Further let $n_k=|\bm{x}^{(k)}|$ and $t_k= \min\big\{\big|\{k'\leq k: p_{k'} \in \mathcal{I}\}\big|- \big|\{k'\leq k: p_{k'} \in \mathcal{J}\}\big|, n_k\big\}$, it can be easily checked that $\bm{x}_{[t_k+1,n_k]}^{(k)}= \bm{y}_{[1,n_k-t_k]}^{(k)}$.
      Now, for $\bm{z}\in \{\bm{x},\bm{y}\}$, we reset $\bm{z}^{(1)} \leftarrow \bm{z}_{[1,p_1-1]} \bm{z}^{(1)}$ and $\bm{z}^{(2t-1)} \leftarrow \bm{z}^{(2t-1)} \bm{z}_{[p_{2t}+1,n]}$.
      It follows that $\bm{x}=\bm{x}^{(1)}\bm{x}^{(2)}\cdots \bm{x}^{(2t-1)}$ and $\bm{y}=\bm{y}^{(1)}\bm{y}^{(2)}\cdots \bm{y}^{(2t-1)}$, where $|\bm{x}^{(k)}|=|\bm{y}^{(k)}|$ and $\mathcal{D}_{t}(\bm{x}^{(k)})\cap \mathcal{D}_{t}(\bm{y}^{(k)})\neq \emptyset$ for $k\in [1,2t-1]$, see Figure \ref{fig:deletion} for an example.
    \end{itemize}
    In both cases, the conclusion is valid for $t=t'$.
    This completes the proof.
\end{IEEEproof}

\begin{lemma}\label{lem:DS}
  For any integer $ t \geq 1 $, let $ \bm{x}, \bm{z}$ be sequences of length $n$ such that $ \mathcal{D}_{t}(\bm{x}) \cap \mathcal{D}_{t}(\bm{z}) \neq \emptyset $. If $ \bm{y} $ can be obtained from $ \bm{z} $ after a substitution at position $ p_y \in [1, n] $, then $ \bm{x} $ and $ \bm{y} $ can be partitioned into $ \bm{x} = \bm{x}^{(1)} \bm{x}^{(2)} $ and $ \bm{y} = \bm{y}^{(1)} \bm{y}^{(2)} $, such that for each $i\in \{1,2\}$, $ |\bm{x}^{(i)}| = |\bm{y}^{(i)}|$ and $ \mathcal{D}_{t}(\bm{x}^{(i)}) \cap \mathcal{D}_{t}(\bm{y}^{(i)}) \neq \emptyset$.
\end{lemma}

\begin{IEEEproof}
Since $\mathcal{D}_{t}(\bm{x})\cap \mathcal{D}_{t}(\bm{z})\neq \emptyset$, there exists some $t'\in [1,t]$ and $p_x,p_z\in [1,n-t'+1]$ such that $\bm{x}_{[1,n]\setminus [p_x,p_x+t'-1]}= \bm{z}_{[1,n]\setminus [p_z,p_z+t'-1]}$.
By symmetry, without loss of generality assume $p_x\leq p_z$.
We distinguish between three cases.
\begin{itemize}
    \item If $p_y< p_x$, we have $\bm{x}_{[1,p_y]\setminus \{p_y\}}= \bm{y}_{[1,p_y]\setminus \{p_y\}}$ and $\bm{x}_{[p_y+1,n]\setminus [p_x,p_x+t'-1]}= \bm{y}_{[p_y+1,n]\setminus [p_z,p_z+t'-1]}$.

    \item If $p_x\leq p_y\leq p_z-t'$, we have $\bm{x}_{[1,p_y+t'-1]\setminus [p_x,p_x+t'-1]}= \bm{y}_{[1,p_y+t'-1]\setminus [p_y,p_y+t'-1]}$ and $\bm{x}_{[p_y+t',n]\setminus [p_y+t',p_y+2t'-1]}= \bm{y}_{[p_y+t',n]\setminus [p_z,p_z+t'-1]}$.
        
    \item If $p_y\geq p_z-t'+1$, we have $\bm{x}_{[1,p_z+t'-1]\setminus [p_x,p_x+t'-1]}= \bm{y}_{[1,p_z+t'-1]\setminus [p_z,p_z+t'-1]}$ and $\bm{x}_{[p_z+t',n]\setminus \{p_y\}}= \bm{y}_{[p_z+t',n]\setminus \{p_y\}}$.
\end{itemize}
Then the conclusion follows.
\end{IEEEproof}

We are now ready to prove Theorem \ref{thm:DS}.

\begin{IEEEproof}[Proof of Theorem \ref{thm:DS}]
    Since the order of insertions, deletions, and substitutions is commutative with respect to the final outcome, the sequence $\bm{y}$ can be obtained from $ \bm{x} $ after $ t $ deletions, then $t$ insertions, followed by $ s'\leq 2s $ substitutions.
    We prove the theorem by induction on $s'$.
    
    For the base case where $s'=0$, the conclusion follows by Lemma \ref{lem:D}.
    Now, assuming that the conclusion is valid for $s'<2s$, we examine the scenario where $s'=2s$.
    Observe that there exists some sequence $\bm{z}$ obtainable from $ \bm{x} $ through the same deletions and insertions, but with $ s'-1 $ substitutions, such that $ \bm{y} $ can be obtained from $ \bm{z} $ by a single substitution, by applying the induction hypothesis on the pair $(\bm{x},\bm{z})$ and using Lemma \ref{lem:DS}, we can conclude that the conclusion is also valid for $s'=2s$.
    This completes the proof.
\end{IEEEproof}

\begin{remark}
  The partitioning technique may be applicable to a broader class of error types.
  For example, in binary alphabets, when the error involves $ t $ deletions combined with $ s $ substitutions and adjacent transpositions, where an adjacent transposition swaps two neighboring symbols, any pair of sequences whose error balls overlap can be partitioned into at most $2t+2s-1$ sub-pairs, where the $^{\leq}t$-burst-deletion balls of each sub-pair intersect. 
  We provide a detailed discussion in the appendix.
  As a result, in the rest of this paper, whenever we construct binary codes capable of correcting $t$ deletions and $s$ substitutions, these codes will also be able to correct $ t $ deletions combined with $ s $ substitutions and adjacent transpositions.
\end{remark}

\section{Code Constructions}\label{sec:code}

We begin by introducing several tools to be used in our constructions. 

\subsection{Sign-Preserving Number and VT Syndromes}

We formalize the concept of the sign-preserving number of an integer sequence.

\begin{definition}
Let $\bm{z}\in \mathbb{Z}^n$, we define its \emph{sign-preserving number}, denoted by $\sigma(\bm{z})$, as the smallest positive integer $k$ such that $\bm{z}$ can be partitioned into $k$ substrings, with the property that within each substring, all non-zero entries share the same sign.
\end{definition}  

It can be easily verified, based on the definition, that the sign-preserving number exhibits the property of subadditivity, as summarized below.

\begin{lemma}\label{lem:number}  
  Let $\bm{z}\in \mathbb{Z}^n$, for $i \in [1,n-1]$, it holds that $\sigma(\bm{z}) \leq \sigma(\bm{z}_{[1,i]}) + \sigma(\bm{z}_{[i+1,n]})$.
  Moreover, if $\bm{z}'$ is a subsequence of $\bm{z}$, then $\sigma(\bm{z}')\leq \sigma(\bm{z})$.  
\end{lemma}  

VT syndromes are typically essential tools in designing codes that can correct deletion-related errors.

\begin{definition}\label{def:VT}
    Let $\bm{z}\in \mathbb{Z}^n$, for any integer $k\geq 0$, the \emph{$k$-th order VT syndrome} of $\bm{z}$ is defined as $\mathrm{VT}^k(\bm{z})=\sum_{i=1}^n i^k x_i$.
\end{definition}

The following lemma is implicit in \cite{Sima-21-IT-kD} and further articulated in \cite{Pi-24-arXiv-2E}.  

\begin{lemma}\cite[Lemma 1]{Pi-24-arXiv-2E}\label{lem:sign}  
  Let $\bm{z}\in \mathbb{Z}^n$ be such that $\mathrm{VT}^k(\bm{z}) = 0$ for $k \in [0,\sigma(\bm{z}) - 1]$, then $\bm{z} = 0^n$.  
\end{lemma}  

\begin{remark}\label{rmk:VT}
  For $ k \geq 1 $, several works define the $ k $-th order VT syndrome of $\bm{z}$ as $\sum_{i=1}^n \sum_{j=1}^i j^{k-1} z_i$.
  The difference between this definition and the one in Definition \ref{def:VT} is minimal when applied to error correction, as Lemma \ref{lem:sign} may still hold for these VT syndromes.
\end{remark}

Lemma \ref{lem:sign} suggests a potential approach to use VT syndromes for designing error-correcting codes. 

\begin{theorem}\label{thm:code}
  For an injection mapping $\bm{\phi}: \Sigma_q^n \rightarrow \mathbb{Z}^{n'}$, where $n$ and $n'$ are positive integers, let $\mathcal{C}\subseteq \Sigma_q^n$ be such that for any $\bm{x},\bm{y}\in \mathcal{C}$ with $\mathcal{B}_{t,s}(\bm{x})\cap \mathcal{B}_{t,s}(\bm{y})\neq \emptyset$, the following holds
  \begin{itemize}
    \item $\sigma\big(\bm{\phi}(\bm{x})-\bm{\phi}(\bm{y})\big)\leq m$;
    \item $\mathrm{VT}^k\big(\bm{\phi}(\bm{x})-\bm{\phi}(\bm{y})\big)=0$ for $k\in [0,m-1]$.
  \end{itemize}
  Then $\bm{x}=\bm{y}$ and $\mathcal{C}$ is a $t$-deletion $s$-substitution correcting code.
\end{theorem}

\begin{IEEEproof}
    Assume $\bm{x}\neq \bm{y}$, since $\bm{\phi}(\cdot)$ is an injection, we have $\bm{\phi}(\bm{x})\neq \bm{\phi}(\bm{y})$.
    Moreover, since $\sigma\big(\bm{\phi}(\bm{x})-\bm{\phi}(\bm{y})\big)\leq m$ and $\mathrm{VT}^k\big(\bm{\phi}(\bm{x})-\bm{\phi}(\bm{y})\big)=0$ for $k\in [0,m-1]$, by Lemma \ref{lem:sign}, we get $\bm{\phi}(\bm{x})= \bm{\phi}(\bm{y})$, which leads to a contradiction.
    Thus, $\bm{x}=\bm{y}$.
    Then, $\mathcal{C}$ is a $t$-deletion $s$-substitution correcting code.
    This completes the proof.
\end{IEEEproof}

\begin{remark}
  Theorem \ref{thm:code} can be adapted to correct a broader class of error types.
\end{remark}

In the following, we identify suitable functions to develop low-redundancy $t$-deletion $s$-substitution correcting codes. Based on Theorem \ref{thm:DS}, the chosen function should be effective in correcting a burst of at most $t$ deletions.
The following functions are known to efficiently correct a burst of at most $t$ deletions.
\begin{itemize}
  \item \emph{Identity Function:} Levenshtein \cite{Levenshtein-66-SPD-1D} utilized it to correct a single deletion in binary alphabets.
  \item \emph{Differential Function:} Levenshtein \cite{Levenshtein-70-BD} employed it to correct a burst of at most two deletions in binary alphabets, while Nguyen et al. \cite{Nguyen-24-IT-1D} used it to correct a single deletion in non-binary alphabets.
\end{itemize}
For other cases, there are no straightforward functions known in the literature that can efficiently correct a burst of at most $t$ deletions.
Therefore, in our code constructions, we focus on the case where $t\leq 2$ for binary alphabets and the case where $t=1$ for non-binary alphabets.
Specifically, we will employ a modified identity function to construct binary codes capable of correcting one deletion and multiple substitutions. Additionally, we will utilize a modified differential function to develop $q$-ary codes that can correct one deletion and multiple substitutions, as well as binary codes for correcting two deletions and multiple substitutions.

\subsection{Accumulative and Accumulative Differential Sequences}

We now introduce the accumulative forms of the identity and differential functions to adapt to our framework.

\begin{definition}
For any $\bm{x} \in \Sigma_q^n$, we define its \emph{accumulative sequence} as $\bm{f}(\bm{x}) = \big(f(\bm{x})_1, f(\bm{x})_2, \ldots, f(\bm{x})_{n}\big)$, where $f(\bm{x})_i = \sum_{j=1}^i x_j $ for $1 \leq i \leq n$.  
\end{definition}

\begin{remark}\label{rmk:f}
  For $k\geq 0$, we have
    \begin{align*}
      \mathrm{VT}^k\big(\bm{f}(\bm{x})\big)
      &= \sum_{i=1}^n i^k f(\bm{x})_i
      = \sum_{i=1}^n i^k \sum_{j=1}^i x_j
      = \sum_{i=1}^n \left(\sum_{j=i}^n j^k \right) x_i
      = \sum_{i=1}^n \left(\sum_{j=1}^{n-i+1} j^k \right) x_{n-i+1}.
    \end{align*}
    Since there is no difference between encoding a sequence and its reversal, Remark \ref{rmk:VT} implies that applying VT syndromes to accumulative sequences is, in essence, equivalent to applying them directly to the original sequences. 
\end{remark}

It can be easily verified that $\bm{f}(\cdot)$ is an injection. For any $i \in [1, n-1]$ and $j\in [i+1,n]$, it holds that
\begin{equation}\label{eq:f}
f(\bm{x})_j = f(\bm{x})_i + f(\bm{x}_{[i+1,n]})_{j-i}.
\end{equation}

\begin{definition}  
For any $\bm{x} \in \Sigma_q^n$, we define its \emph{differential sequence} as $\bm{d}(\bm{x}) = \big(d(\bm{x})_1, d(\bm{x})_2, \ldots, d(\bm{x})_{n}\big)$ and its \emph{accumulative differential sequence} as $\bm{g}(\bm{x}) = \big(g(\bm{x})_1, g(\bm{x})_2, \ldots,g(\bm{x})_{n}\big)$, where $d(\bm{x})_i = x_i - x_{i-1} \pmod{q}$ and $g(\bm{x})_i = \sum_{j=1}^i d(\bm{x})_j $ for $i \in [1,n]$.  
\end{definition}

\begin{remark}\label{rmk:run}
  When $q=2$, the accumulative differential sequence corresponds to the run string used in \cite{Levenshtein-70-BD, Guruswami-21-IT-2D, Gabrys-23-IT-DS, Pi-24-arXiv-2E}.
\end{remark}

\begin{remark}\label{rmk:g}
  Similar to the discussion in Remark \ref{rmk:f}, we can conclude that applying VT syndromes to accumulative differential sequences is, in essence, equivalent to applying them directly to differential sequences.
\end{remark}

It can be easily verified that $\bm{b}(\cdot)$ is a bijection and that $\bm{g}(\cdot)$ is an injection. 
For any $i \in [1, n]$, we have
\begin{equation}\label{eq:g'}
  g(\bm{x})_i\equiv x_i \pmod{q}.
\end{equation}
Moreover, for $j\in [i+1,n]$, it holds that
\begin{equation}\label{eq:g}
  g(\bm{x})_j= g(\bm{x})_i+ g(\bm{x}_{[i,n]})_{j-i+1}-x_i.
\end{equation}

\begin{lemma}\label{lem:insertion}  
   Let $\bm{x},\bm{y}\in \Sigma_q^n$ and $p\in [1,n]$, we define $\bm{x}'=\bm{x}_{[1,p]}x_{p+1}y_p\bm{x}_{[p+1,n]}$ and $\bm{y}'=\bm{y}_{[1,p]}x_{p+1}y_p\bm{y}_{[p+1,n]}$, then $\sigma\big(\bm{g}(\bm{x}) - \bm{g}(\bm{y})\big) \leq \sigma\big(\bm{g}(\bm{x}') - \bm{g}(\bm{y}')\big)$.
\end{lemma}

\begin{IEEEproof}
    Observe that $g(\bm{x}')_i-g(\bm{y}')_i= g(\bm{x})_i-g(\bm{y})_i$ for $i\leq p$.
    Moreover, by Equation (\ref{eq:g}), we can compute $g(\bm{x}')_i-g(\bm{y}')_i= g(\bm{x})_{i-2}-g(\bm{y})_{i-2}$ for $i\geq p+3$.
    This implies that $\bm{g}(\bm{x}) - \bm{g}(\bm{y})$ is a subsequence of $\bm{g}(\bm{x}') - \bm{g}(\bm{y}')$.
    Then by Lemma \ref{lem:number}, we get $\sigma\big(\bm{g}(\bm{x}) - \bm{g}(\bm{y})\big) \leq \sigma\big(\bm{g}(\bm{x}') - \bm{g}(\bm{y}')\big)$.
\end{IEEEproof}

Since, later in our code constructions, VT syndromes are applied to accumulative sequences and accumulative differential sequences rather than directly to the original sequences, we first examine how errors affect these sequences. 
By Theorem \ref{thm:DS}, it suffices to consider the burst-deletion error.

\subsubsection{The Influence of Burst-Deletion on Accumulative Sequences}

\begin{lemma}\label{lem:del}
    Let $\bm{x}, \bm{y} \in \Sigma_2^n$ be such that $\bm{x}_{[1,n] \setminus \{p_x\}} = \bm{y}_{[1,n] \setminus \{p_y\}}$ for some $1\leq p_x \leq p_y \leq n$, we have either $f(\bm{x})_{i}-f(\bm{y})_{i}\in \{0,1\}$ for $i\in [1,n]$ or $f(\bm{x})_{i}-f(\bm{y})_{i}\in \{-1,0\}$ for $i\in [1,n]$.
   For $k\geq 0$, it follows that
      \begin{equation}\label{eq:VT_del}
        \left| \mathrm{VT}^k\big(\bm{f}(\bm{x})\big)- \mathrm{VT}^k\big(\bm{f}(\bm{y})\big) \right|\leq \sum_{i=1}^n i^k.
      \end{equation}
\end{lemma}

\begin{figure*}
\center
\begin{align*}
    \xymatrix@C=1ex@R=1ex{
    \bm{x}: & \ast \ar@{-}[d] & \cdots \ar@{-}[d] & \ast \ar@{-}[d] & p_x & \cdots & p_{x}+t-1 & \ast \ar@{-}[dlll] & \cdots \ar@{-}[dlll] & \ast \ar@{-}[dlll] & \ast \ar@{-}[d] & \cdots \ar@{-}[d] & \ast \ar@{-}[d]\\
    \bm{y}: & \ast & \cdots & \ast & \ast & \cdots & \ast & p_y & \cdots & p_y+t-1 & \ast & \cdots & \ast
    }
\end{align*}
\caption{Illustrations of $\bm{x}$ and $\bm{y}$ when $\bm{x}_{[1,n] \setminus [p_x,p_x+t-1]} = \bm{y}_{[1,n] \setminus [p_y+t-1]}$, where $t\geq 1$ and $1\leq p_x\leq p_y\leq n-t+1$. Each pair of bits connected by a solid line are of equal value.}
\label{fig:burst-deletion}
\end{figure*}
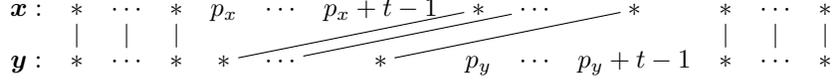
    
\begin{IEEEproof}
  By Figure \ref{fig:burst-deletion} and the definition of the accumulative sequence, we can compute 
  \begin{equation}\label{eq:identity}
      f(\bm{x})_{i}-f(\bm{y})_{i}=
      \begin{cases}
          0, &\mbox{if } i\in [0,p_x-1];\\
          x_{p_x}- y_i, &\mbox{if } i \in [p_x, p_y];\\
          x_{p_x}- y_{p_y}, &\mbox{if } i \in [p_y+1,n].
      \end{cases}
  \end{equation}
  This implies that $f(\bm{x})_{i}-f(\bm{y})_{i}\in \{0,1\}$ for $i\in [1,n]$ if $x_{p_x}=1$, otherwise $f(\bm{x})_{i}-f(\bm{y})_{i}\in \{0,-1\}$ for $i\in [1,n]$.
\end{IEEEproof}  

\begin{remark}\label{rmk:AS}
   By Lemma \ref{lem:del}, for any sequences $ \bm{x}, \bm{y} $ with non-empty intersection of their single-deletion balls, it holds that $\sigma\big(\bm{f}(\bm{x}) - \bm{f}(\bm{y}) \big)\leq 1$.
   Then by Theorem \ref{thm:DS}, it may be suitable to apply VT syndromes on accumulative sequences to construct binary single-deletion $s$-substitution correcting codes. 
   However, this approach encounters limitations when correcting multiple deletions in binary alphabets or even a single deletion in non-binary alphabets. 
   We illustrate these limitations as follows.
    \begin{itemize}
      \item \emph{Non-binary alphabet:} For any $m\leq \frac{n-1}{2}$ and $\bm{z}\in \Sigma_q^{n-2m-1}$, we consider the sequences $\bm{x}= \big(1,(q-1,0)^m,\bm{z}\big)$ and $\bm{y}= \big((q-1,0)^m,\bm{z},1\big)$. 
          It can be easily verified that $\bm{x}_{[2,n]}= \bm{y}_{[1,n-1]}$.
          Furthermore, for $i\in [1,2m]$, we have
          \begin{equation*}
            f(\bm{x})_i-f(\bm{y})_i=
            \begin{cases}
              2-q<0, & \mbox{if } i\equiv 1 \pmod{2};\\
              1, & \mbox{if } i\equiv 0 \pmod{2}.
            \end{cases}
          \end{equation*}
          This implies that $\sigma\big(\bm{f}(\bm{x})-\bm{f}(\bm{y}) \big)\geq 2m$.
          Consequently, this method becomes inefficient even for correcting a single deletion. 
          The core reason is that, in Equation (\ref{eq:identity}), the sign of $ x_{p_x} - y_i $ (regardless of $ y_i $) depends solely on $ x_{p_x} $ in binary alphabets, but not in non-binary ones.
      \item \emph{Binary alphabet:} For any $m\leq \frac{n-2}{4}$ and $\bm{z}\in \Sigma_2^{n-4m-2}$, we consider the sequences $\bm{x}= \big(1,0,(1,1,0,0)^m,\bm{z}\big)$ and $\bm{y}= \big((1,1,0,0)^m,\bm{z},1,0\big)$. 
          It can be easily verified that $ \bm{x}_{[3, n]} = \bm{y}_{[1, n-2]} $. Furthermore, for $i\in [1,4m]$, we have
          \begin{equation*}
            f(\bm{x})_i-f(\bm{y})_i=
            \begin{cases}
              0, & \mbox{if } i\equiv 1 \pmod{2};\\
              -1, & \mbox{if } i\equiv 2 \pmod{4};\\
              1, & \mbox{if } i\equiv 0 \pmod{4}.
            \end{cases}
          \end{equation*}
          This implies that $\sigma\big(\bm{f}(\bm{x})-\bm{f}(\bm{y}) \big)\geq 2m$.
          Thus, the approach becomes inefficient for correcting multiple deletions.
          The limited effectiveness here stems from the fact that the sign of $ \sum_{i=1}^t x_i - \sum_{i=1}^t y_i $ can not be determined solely by $ \sum_{i=1}^t x_i $ when $ t\geq 2 $.
          In Section \ref{subsec:binary}, we will present a strategy to address this challenge.
      \end{itemize}
      By Remark \ref{rmk:f}, the approach of applying VT syndromes directly to original sequences may inherit similar limitations.
\end{remark}

\subsubsection{The Influence of Burst-Deletion on Accumulative Differential Sequences}

\begin{lemma}\label{lem:del'}  
  Let $\bm{x}, \bm{y} \in \Sigma_q^n$ be such that $\bm{x}_{[1,n] \setminus \{p_x\}} = \bm{y}_{[1,n] \setminus \{p_y\}}$ for some $1\leq p_x\leq p_y \leq n$, we have either $g(\bm{x})_{i}-g(\bm{y})_{i}\in [0,q]$ for $i\in [1,n]$ or $g(\bm{x})_{i}-g(\bm{y})_{i}\in [-q,0]$ for $i\in [1,n]$.
  Moreover, it is clear that $|g(\bm{x})_{1}-g(\bm{y})_{1}|\leq q-1$.
  For $k\geq 0$, it follows that
  \begin{equation}\label{eq:VT_del'}
    \left| \mathrm{VT}^k\big(\bm{g}(\bm{x})\big)- \mathrm{VT}^k\big(\bm{g}(\bm{y})\big) \right| \leq q\sum_{i=1}^{n} i^k-1.
  \end{equation}
\end{lemma}

\begin{figure*}
\center
\begin{align*}
    \xymatrix@C=1ex@R=1ex{
    \bm{d}(\bm{x}): & \ast \ar@{-}[d] & \cdots \ar@{-}[d] & \ast \ar@{-}[d] & p_x & \cdots & p_{x}+t & \ast \ar@{-}[dll] & \cdots \ar@{-}[dll] & \ast \ar@{-}[dll] & \ast & \ast \ar@{-}[d] & \cdots \ar@{-}[d] & \ast \ar@{-}[d]\\
    \bm{d}(\bm{y}): & \ast & \cdots & \ast & \ast & \ast & \cdots & \ast & p_y & \cdots & p_y+t & \ast & \cdots & \ast
    }
\end{align*}
\caption{Illustrations of $\bm{d}(\bm{x})$ and $\bm{d}(\bm{y})$ when $\bm{x}_{[1,n] \setminus [p_x,p_x+t-1]} = \bm{y}_{[1,n] \setminus [p_y+t-1]}$, where $t\geq 1$ and $1\leq p_x< p_y\leq n-t+1$. Each pair of bits connected by a solid line are of equal value.}
\label{fig:burst-deletion'}
\end{figure*}
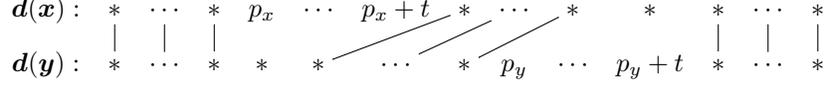

\begin{IEEEproof}
  We first consider the case where $p_x<p_y$.
  By Figure \ref{fig:burst-deletion} and the definition of the differential sequence, we get $d(\bm{x})_{p_x}+ d(\bm{x})_{p_x+1}\equiv d(\bm{y})_{d_x} \pmod{q}$ and $d(\bm{x})_{d_y+1}\equiv d(\bm{y})_{p_y}+ d(\bm{y})_{p_y+1} \pmod{q}$.
  This implies that
  \begin{gather}\label{eq:property}
      d(\bm{x})_{p_x}+ d(\bm{x})_{p_x+1}- d(\bm{y})_{d_x}\in \{0,q\},\\
      d(\bm{x})_{d_y+1}- d(\bm{y})_{p_y}- d(\bm{y})_{p_y+1}\in \{-q,0\} \nonumber.
  \end{gather}
  By Figure \ref{fig:burst-deletion'} and the definition of the accumulative differential sequence, we can compute
  \begin{equation*}
      g(\bm{x})_{i}- g(\bm{y})_{i}=
      \begin{cases}
          0, &\mbox{if } i\in [1,p_x-1];\\
          d(\bm{x})_{p_x}- d(\bm{y})_{p_x}, &\mbox{if } i=p_x;\\
          d(\bm{x})_{p_x}+ d(\bm{x})_{p_x+1}- d(\bm{y})_{p_x}-d(\bm{y})_i, &\mbox{if } i\in [p_x+1,p_y];\\
          d(\bm{x})_{p_x}+ d(\bm{x})_{p_x+1}- d(\bm{y})_{p_x}+ d(\bm{x})_{p_y+1}- d(\bm{y})_{p_y}- d(\bm{y})_{p_y+1}, &\mbox{if } i\in [p_y+1,n].
      \end{cases}
  \end{equation*}
  This implies that $g(\bm{x})_{i}-g(\bm{y})_{i}\in [0,q]$ for $i\in [1,n]$ if $d(\bm{x})_{p_x}+ d(\bm{x})_{p_x+1}- d(\bm{y})_{d_x}=q$, otherwise $g(\bm{x})_{i}-g(\bm{y})_{i}\in [-q,0]$ for $i\in [1,n]$.
  Thus, the conclusion holds for $p_x<p_y$.
  
  For the case where $p_x=p_y$, we consider the sequences $\bm{x}'=\bm{x}_{[1,p_x]} x_{p_x+1} y_{p_x} \bm{x}_{[p_x+1,n]}$ and $\bm{y}'=\bm{y}_{[1,p_x]} x_{p_x+1} y_{p_x} \bm{y}_{[p_x+1,n]}$.
  By Lemma \ref{lem:insertion}, $\bm{g}(\bm{x})-\bm{g}(\bm{y})$ is a subsequence of $\bm{g}(\bm{x}')-\bm{g}(\bm{y}')$.
  Observe that $\bm{x}_{[1,n+2]\setminus \{p_x\}}'= \bm{y}_{[1,n+2]\setminus \{p_x+2\}}'$, by applying the result from the previous case, we can conclude that the statement also holds for $p_x=p_y$.
  This completes the proof.
\end{IEEEproof}


\begin{lemma}\label{lem:burst-del}  
  Let $\bm{x}, \bm{y} \in \Sigma_2^n$ be such that $\bm{x}_{[1,n] \setminus \{p_x,p_x+1\}} = \bm{y}_{[1,n] \setminus \{p_y,p_y+1\}}$ for some $1\leq p_x\leq p_y \leq n-1$, we have either $g(\bm{x})_{i}-g(\bm{y})_{i}\in [0,2]$ for $i\in [1,n]$ or $g(\bm{x})_{i}-g(\bm{y})_{i}\in [-2,0]$ for $i\in [1,n]$.
  Moreover, it is clear that $|g(\bm{x})_{1}-g(\bm{y})_{1}|\leq 1$.
  For $k\geq 0$, it follows that
  \begin{equation}\label{eq:burst'}
    \left| \mathrm{VT}^k\big(\bm{g}(\bm{x})\big)- \mathrm{VT}^k\big(\bm{g}(\bm{y})\big) \right|\leq 2\sum_{i=1}^{n} i^k-1.
  \end{equation}
\end{lemma}

\begin{IEEEproof}
  We first consider the case where $p_x<p_y$.
  By Figure \ref{fig:burst-deletion} and the definition of the differential sequence, we get $d(\bm{x})_{p_x}+ d(\bm{x})_{p_x+1}+d(\bm{x})_{p_x+2}\equiv d(\bm{y})_{d_x} \pmod{2}$ and $d(\bm{x})_{p_y+2}\equiv d(\bm{y})_{p_y}+ d(\bm{y})_{p_y+1}+d(\bm{y})_{p_y+2} \pmod{2}$.
  This implies that
  \begin{gather}\label{eq:property'}
      d(\bm{x})_{p_x}+ d(\bm{x})_{p_x+1}+ d(\bm{x})_{p_x+2}- d(\bm{y})_{p_x}\in \{0,2\},\\
      d(\bm{x})_{p_y+2}- d(\bm{y})_{p_y}- d(\bm{y})_{p_y+1}- d(\bm{y})_{p_y+2}\in \{-2,0\}\nonumber.
  \end{gather}
  By Figure \ref{fig:burst-deletion'} and the definition of the accumulative differential sequence, we can compute
  \begin{equation}\label{eq:differential}
      g(\bm{x})_{i}- g(\bm{y})_{i}=
      \begin{cases}
          0, &\mbox{if } i\in [1,p_x-1];\\
          d(\bm{x})_{p_x}- d(\bm{y})_{p_x}, &\mbox{if } i=p_x;\\
          d(\bm{x})_{p_x}+ d(\bm{x})_{p_x+1}- d(\bm{y})_{p_x}-d(\bm{y})_{p_x+1}, &\mbox{if } i=p_x+1;\\
          d(\bm{x})_{p_x}+ d(\bm{x})_{p_x+1}+ d(\bm{x})_{p_x+2}- d(\bm{y})_{p_x}-d(\bm{y})_{i-1}-d(\bm{y})_i, &\mbox{if } i\in [p_x+2,p_y+1];\\
          d(\bm{x})_{p_x}+ d(\bm{x})_{p_x+1}+ d(\bm{x})_{p_x+2}- d(\bm{y})_{p_x}\\
          \quad\quad\quad + d(\bm{x})_{d_y+2}- d(\bm{y})_{p_y}- d(\bm{y})_{p_y+1}- d(\bm{y})_{p_y+2}, &\mbox{if } i\in [p_y+2,n].
      \end{cases}
  \end{equation}
  This implies that $g(\bm{x})_{i}-g(\bm{y})_{i}\in [0,2]$ for $i\in [1,n]$ if $d(\bm{x})_{p_x}+ d(\bm{x})_{p_x+1}+d(\bm{x})_{p_x+2}- d(\bm{y})_{d_x}=2$, otherwise $g(\bm{x})_{i}-g(\bm{y})_{i}\in [-2,0]$ for $i\in [1,n]$.
  Thus, the conclusion holds for $p_x<p_y$.
  
  For the case where $p_x=p_y$, we consider the sequences $\bm{x}'=\bm{x}_{[1,p_x]} x_{p_x+1} y_{p_x} \bm{x}_{[p_x+1,n]}$ and $\bm{y}'=\bm{y}_{[1,p_x]} x_{p_x+1} y_{p_x} \bm{y}_{[p_x+1,n]}$.
  By Lemma \ref{lem:insertion}, $\bm{g}(\bm{x})-\bm{g}(\bm{y})$ is a subsequence of $\bm{g}(\bm{x}')-\bm{g}(\bm{y}')$.
  Observe that $\bm{x}_{[1,n+2]\setminus \{p_x,p_x+1\}}'= \bm{y}_{[1,n+2]\setminus \{p_x+2,p_x+3\}}'$, by applying the result from the previous case, we can conclude that the statement also holds for $p_x=p_y$.
  This completes the proof.
\end{IEEEproof}

\begin{remark}\label{rmk:ADS}
  By Lemma \ref{lem:del'}, for any sequences $ \bm{x}, \bm{y} $ with non-empty intersection of their single-deletion balls, it holds that $\sigma\big(\bm{g}(\bm{x}) - \bm{g}(\bm{y}) \big)\leq 1$.
  Then by Theorem \ref{thm:DS}, it may be suitable to apply VT syndromes on accumulative differential sequences to construct non-binary single-deletion $s$-substitution correcting codes. 
  Similarly, by Theorem \ref{thm:DS} and Lemmas \ref{lem:del'} and \ref{lem:burst-del}, this approach may be also valid to construct binary two-deletion $s$-substitution correcting codes.
  The intuition that differential sequences are more effective than the sequences themselves stems from the properties expressed in Equations (\ref{eq:property}) and (\ref{eq:property'}).
  However, this approach encounters limitations when correcting three deletions in binary alphabets or even two deletions in non-binary alphabets. 
  We illustrate these limitations as follows.
  \begin{itemize}
      \item \emph{Non-binary alphabet:} For any $m\leq \frac{n-2}{4}$ and $\bm{z}\in \Sigma_q^{n-4m-2}$, let $\bm{z}'$ be such that $\bm{d}(\bm{z}')=\big((q-1,q-1,0,0)^m,\bm{z}\big)$, we consider the sequences $\bm{x}= \big(1,0,\bm{z}'\big)$ and $\bm{y}= \big(\bm{z}',1,0\big)$. 
          It is straightforward to verify that $\bm{x}_{[3,n]}= \bm{y}_{[1,n-2]}$.
          Moreover, we have $\bm{d}(\bm{x})=\big(1,q-1,(q-1,q-1,0,0)^m,\bm{z}\big)$ and $\bm{d}(\bm{y})_{[1,n-2]}=\big((q-1,q-1,0,0)^m,\bm{z}\big)$.
          Then for $i\in [2,4m]$, it holds that 
          \begin{equation*}
            g(\bm{x})_i-g(\bm{y})_i=
            \begin{cases}
              1, & \mbox{if } i\equiv 1 \pmod{2};\\
              2-q<0, & \mbox{if } i\equiv 2 \pmod{4};\\
              q, & \mbox{if } i\equiv 0 \pmod{4}.
            \end{cases}
          \end{equation*}
          This implies that $\sigma\big(\bm{g}(\bm{x})-\bm{g}(\bm{y}) \big)\geq 2m$.
          Consequently, this approach becomes inefficient for correcting multiple deletions.
          The key reason for the difference between binary and non-binary alphabets is that, in Equation (\ref{eq:differential}), the sign of $d(\bm{x})_{p_x}+ d(\bm{x})_{p_x+1}+ d(\bm{x})_{p_x+2}- d(\bm{y})_{p_x}-d(\bm{y})_{j-1}-d(\bm{y})_j$ (regardless of $d(\bm{y})_{j-1}+d(\bm{y})_j$), where $d(\bm{x})_{p_x} + d(\bm{x})_{p_x+1} + d(\bm{x})_{p_x+2} - d(\bm{y})_{p_x} \in \{0, q\}$, can be determined solely by $d(\bm{x})_{p_x} + d(\bm{x})_{p_x+1} + d(\bm{x})_{p_x+2} - d(\bm{y})_{p_x}$ in binary alphabets, but not in non-binary ones.
      \item \emph{Binary alphabet:} For any $m\geq \frac{n-3}{6}$ and $\bm{z}\in \Sigma_2^{n-6m-3}$, let $\bm{z}'$ be such that $\bm{d}(\bm{z})=\big((1,1,1,0,0,0)^m,\bm{z}'\big)$, we consider the sequences $\bm{x}= \big(1,1,0,\bm{z}'\big)$ and $\bm{y}= \big(\bm{z}',1,1,0\big)$. 
          It is straightforward to verify that $\bm{x}_{[4,n]}= \bm{y}_{[1,n-3]}$. Moreover, we have $\bm{d}(\bm{x})=\big(1,0,1,(1,1,1,0,0,0)^m,\bm{z}\big)$ and $\bm{d}(\bm{y})_{[1,n-3]}=\big((1,1,1,0,0,0)^m,\bm{z}\big)$.
          Then for $i\in [3,6m]$, it holds that 
          \begin{equation*}
            g(\bm{x})_i-g(\bm{y})_i=
            \begin{cases}
              -1, & \mbox{if } i\equiv 3 \pmod{6};\\
              0, & \mbox{if } i\equiv 2 \pmod{6} \text{ or } i\equiv 4 \pmod{6};\\
              1, & \mbox{if } i\equiv 1 \pmod{6} \text{ or } i\equiv 5 \pmod{6};\\
              2, & \mbox{if } i\equiv 0 \pmod{6}.
            \end{cases}
          \end{equation*}
          This implies that $\sigma\big(\bm{g}(\bm{x})-\bm{g}(\bm{y}) \big)\geq 2m$.
          Consequently, this approach becomes inefficient for correcting three deletions.
    \end{itemize}
    By Remark \ref{rmk:g}, the approach of simply applying VT syndromes to differential sequences may inherit similar limitations.
\end{remark}

\subsection{Code Constructions}

In this subsection, we introduce our code constructions for correcting deletions and substitutions.

\subsubsection{Constructions Using Accumulative Sequences}

As explained in Remark \ref{rmk:AS}, applying VT syndromes on accumulative sequences may be efficient to construct binary codes for correcting one deletion and multiple substitutions.
In the following, we will use this approach to construct binary single-deletion $s$-substitution correcting codes with a redundancy of at most $(s+1)(2s+1)\log n + O(1)$ bits. 
This is achieved by using the following conclusion. 

\begin{lemma}\label{lem:AS}
  Let $\bm{x},\bm{y}\in \Sigma_2^n$ be such that $\bm{x}=\bm{x}^{(1)}\bm{x}^{(2)}\cdots \bm{x}^{(m)}$ and $\bm{y}=\bm{y}^{(1)}\bm{y}^{(2)}\cdots \bm{y}^{(m)}$, where $m\geq 1$ and  for $i\in [1,m]$, $|\bm{x}^{(i)}|=|\bm{y}^{(i)}|$ and $\mathcal{D}_{1}(\bm{x}^{(i)})\cap \mathcal{D}_{1}(\bm{y}^{(i)})\neq \emptyset$, then 
  \begin{itemize}
    \item $\sigma\big(\bm{f}(\bm{x})-\bm{f}(\bm{y})\big)\leq m$;
    \item $\big| \mathrm{VT}^k\big(\bm{f}(\bm{x})-\bm{f}(\bm{y})\big) \big|\leq m\sum_{i=1}^n i^k$ for $k\geq 0$.
  \end{itemize}
\end{lemma}

\begin{IEEEproof}
    We first show that $\sigma\big(\bm{f}(\bm{x})-\bm{f}(\bm{y})\big)\leq m$.
    For $i\in [1,m]$, let $p_i= \sum_{j=1}^i |\bm{x}^{(i)}|$.
    Further let $p_0=0$.
    By Equation (\ref{eq:f}) and Lemma \ref{lem:del}, for $i\in [0,m-1]$, we can compute $\sigma\big( \bm{f}(\bm{x})_{[p_i+1,p_{i+1}]} - \bm{f}(\bm{y})_{[p_i+1,p_{i+1}]} \big)=1$.
    It then follows by Lemma \ref{lem:number} that
    \begin{align*}
      \sigma\big( \bm{f}(\bm{x})- \bm{f}(\bm{y}) \big)
      &\leq \sum_{i=0}^{m-1} \sigma\big( \bm{f}(\bm{x})_{[p_i+1,p_{i+1}]} - \bm{f}(\bm{y})_{[p_i+1,p_{i+1}]} \big)
      =m.
    \end{align*}
    
    Now we show that $\big| \mathrm{VT}^k\big(\bm{f}(\bm{x})-\bm{f}(\bm{y})\big) \big|\leq m\sum_{i=1}^n i^k$ for $k\geq 0$.
    For $i\in [0,m]$, let $\bm{z}^{(i)}= \bm{y}^{(1)}\cdots \bm{y}^{(i)}\bm{x}^{(i+1)}\cdots \bm{x}^{(m)}$.  
    It follows that $\bm{z}^{(0)}=\bm{x}$, $\bm{z}^{(m)}=\bm{y}$, and $\mathcal{D}_1(\bm{z}^{(i)})\cap \mathcal{D}_1(\bm{z}^{(i+1)})\neq \emptyset$ for $i\in [0,m-1]$.
    Then we can compute
    \begin{align*}  
     \left| \mathrm{VT}^k\big(\bm{f}(\bm{x})-\bm{f}(\bm{y})\big) \right|
     &=\left| \mathrm{VT}^k\big(\bm{f}(\bm{x})\big) - \mathrm{VT}^k\big(\bm{f}(\bm{y})\big) \right|\\
     &=\left| \sum_{i=0}^{m-1}\mathrm{VT}^k\big(\bm{f}(\bm{z}^{(i)})\big) - \mathrm{VT}^k\big(\bm{f}(\bm{z}^{(i+1)})\big) \right|\\
     &\leq \sum_{i=0}^{m-1} \left| \mathrm{VT}^k \big(\bm{f}(\bm{z}^{(i)})\big) - \mathrm{VT}^k\big(\bm{f}(\bm{z}^{(i+1)})\big) \right|\\ 
     &\leq m\sum_{i=1}^n i^k,
    \end{align*} 
    where the last inequality follows by Equation (\ref{eq:VT_del}).
    This completes the proof.
\end{IEEEproof}

Now by Theorems \ref{thm:DS} and \ref{thm:code} and Lemma \ref{lem:AS}, we can derive the following code construction.

\begin{theorem}\label{thm:t=1}
  For $k\in [0,2s]$, let $n_k\triangleq (2s+1)\sum_{i=1}^n i^k$ and $a_k\in [0,n_k]$, the code
  \begin{align*}
    \mathcal{C}_1= \big\{\bm{x}\in \Sigma_2^n: \mathrm{VT}^k\big(\bm{f}(\bm{x})\big) \equiv a_{k} \pmod{n_k+1} \text{ for } k\in [0,2s]\big\}
  \end{align*}
  is a binary single-deletion $s$-substitution correcting code.
  Moreover, by the pigeonhole principle, there exists a choice of parameters such that its redundancy is at most $(s+1)(2s+1)\log n+O(1)$ bits.
\end{theorem}

\begin{IEEEproof}
    Assume there are two sequences $\bm{x},\bm{y} \in \mathcal{C}_1$ such that $\mathcal{B}_{1,s}(\bm{x}) \cap \mathcal{B}_{1,s}(\bm{y})\neq \emptyset$.
    By Theorem \ref{thm:DS}, there exists some $m\in [1,2s+1]$ such that $\bm{x}=\bm{x}^{(1)}\bm{x}^{(2)}\cdots \bm{x}^{(m)}$ and $\bm{y}=\bm{y}^{(1)}\bm{y}^{(2)}\cdots \bm{y}^{(m)}$, where $|\bm{x}^{(i)}|=|\bm{y}^{(i)}|$ and $\mathcal{D}_{1}(\bm{x}^{(i)})\cap \mathcal{D}_{1}(\bm{y}^{(i)})\neq \emptyset$ for $i\in [1,m]$.
    Then by Lemma \ref{lem:AS} and the condition that $\mathrm{VT}^k\big(\bm{f}(\bm{x})\big)\equiv \mathrm{VT}^k\big(\bm{f}(\bm{y})\big) \pmod{n_k+1}$, we have $\sigma\big(\bm{f}(\bm{x})-\bm{f}(\bm{y})\big)\leq m$ and $\mathrm{VT}^k\big(\bm{f}(\bm{x})-\bm{f}(\bm{y})\big)=0$ for $k\in [0,m-1]$.
    Finally, by Theorem \ref{thm:code}, $\mathcal{C}_1$ is indeed a single-deletion $s$-substitution correcting code.
    
    Now, by considering all possible choices for $a_0,a_1,\ldots,a_{2s}$, we observe that the total number of codes $\mathcal{C}_1$ is $\prod_{k=0}^{2s}(n_k+1)$. 
    These codes partition the space $\Sigma_2^{n}$. 
    By the pigeonhole principle, there exists at least one choice of $a_0,a_1,\ldots,a_{2s}$ such that the code $\mathcal{C}_1$ contains at least $\frac{2^{n}}{\prod_{k=0}^{2s}(n_k+1)}= O\big(\frac{2^n}{n^{(s+1)(2s+1)}}\big)$ codewords. 
    The redundancy of such a code is at most $(s+1)(2s+1)\log n + O(1)$ bits.
    This completes the proof.
\end{IEEEproof}

\begin{remark}
    Since by Remark \ref{rmk:f} that applying VT syndromes to accumulative sequences is, in essence, equivalent to applying them directly to the original sequences, the code presented in Theorem \ref{thm:t=1} aligns with the constructions of Levenshtein \cite{Levenshtein-66-SPD-1D} for $s=0$, Smagloy et al. \cite{Smagloy-23-IT-DS} for $s=1$, and Song et al. \cite{Song-22-IT-DS} for $s \geq 1$, respectively. 
    Moreover, compared to \cite{Smagloy-23-IT-DS, Song-22-IT-DS}, our proof is relatively simpler.
\end{remark}

\subsubsection{Constructions Using Accumulative Differential Sequences}

As explained in Remark \ref{rmk:ADS}, applying VT syndromes on accumulative differential sequences may be efficient to construct non-binary codes for correcting one deletion and multiple substitutions, as well as binary codes for correcting two deletions and multiple substitutions.
In the following, we will use this approach to construct non-binary single-deletion $s$-substitution correcting codes with a redundancy of at most $(s+1)(2s+1)\log n + O(1)$ bits, as well as binary two-deletion $s$-substitution correcting codes with a redundancy of at most $(s+1)(2s+3)\log n + O(1)$ bits. This is achieved by using the following conclusion.

\begin{lemma}\label{lem:ADS}
  Let $\bm{x},\bm{y}\in \Sigma_q^n$ be such that $\bm{x}=\bm{x}^{(1)}\bm{x}^{(2)}\cdots \bm{x}^{(m)}$ and $\bm{y}=\bm{y}^{(1)}\bm{y}^{(2)}\cdots \bm{y}^{(m)}$, where $m\geq 1$ and $|\bm{x}^{(i)}|=|\bm{y}^{(i)}|$ for $i\in [1,m]$.
  If $\mathcal{D}_{1}(\bm{x}^{(i)})\cap \mathcal{D}_{1}(\bm{y}^{(i)})\neq \emptyset$ when $q> 2$ and $\mathcal{D}_{2}(\bm{x}^{(i)})\cap \mathcal{D}_{2}(\bm{y}^{(i)})\neq \emptyset$ when $q=2$, the following holds:
  \begin{itemize}
    \item $\sigma\big(\bm{g}(\bm{x})-\bm{g}(\bm{y})\big)\leq m$;
    \item $\big| \mathrm{VT}^k\big(\bm{g}(\bm{x})-\bm{g}(\bm{y})\big) \big|\leq qm\sum_{i=1}^n i^k-m$ for $k\geq 0$.
  \end{itemize}
\end{lemma}

\begin{IEEEproof}
    We first show that $\sigma\big(\bm{g}(\bm{x})-\bm{g}(\bm{y})\big)\leq m$.
    Set $\bm{x}^{(m+1)}=0$.
    For $i\in [1,m]$ and $\bm{v}\in \{\bm{x},\bm{y}\}$, let $\tilde{\bm{v}}^{(i)}= \bm{v}^{(i)} \alpha_i \beta_i$, where $\alpha_i$ is the first entry of $\bm{x}^{(i+1)}$ and $\beta_i$ is the last entry of $\bm{y}^{(i)}$, it follows that $\mathcal{D}_{1}(\tilde{\bm{x}}^{(i)})\cap \mathcal{D}_{1}(\tilde{\bm{y}}^{(i)})\neq \emptyset$ when $q> 2$ and $\mathcal{D}_2(\tilde{\bm{x}}^{(i)})\cap \mathcal{D}_{2}(\tilde{\bm{y}}^{(i)})\neq \emptyset$ when $q=2$.
    Let $\bm{x}'= \tilde{\bm{x}}^{(1)}\tilde{\bm{x}}^{(2)}\cdots \tilde{\bm{x}}^{(m)}$ and $\bm{y}'= \tilde{\bm{y}}^{(1)}\tilde{\bm{y}}^{(2)} \cdots \tilde{\bm{y}}^{(m)}$.
    By Lemma \ref{lem:insertion}, we get $\sigma\big( \bm{g}(\bm{x}) - \bm{g}(\bm{y}) \big)\leq \sigma\big( \bm{g}(\bm{x}') - \bm{g}(\bm{y}') \big)$.
    It thus suffices to show that $\sigma\big( \bm{g}(\bm{x}') - \bm{g}(\bm{y}') \big)\leq m$.
    For $i\in [0,m]$, let $p_i= \sum_{j=1}^i |\tilde{\bm{x}}^{(i)}|$. 
    By Equation (\ref{eq:g}), for $j\in [p_i+1,p_{i+1}]$, we can compute
    \begin{align*}
      g(\bm{x}')_j- g(\bm{y}')_j
      &= g(\bm{x}')_{p_i}+g(\bm{x}_{[p_i,p_{p+1}]}')_{j-p_i+1}- x_{p_i}'- g(\bm{y}')_{p_i}-g(\bm{y}_{[p_i,p_{p+1}]}')_{j-p_i+1}+ y_{p_i}'\\
      &= g(\bm{x}')_{p_i}-g(\bm{y}')_{p_i}+ g(\bm{x}_{[p_i,p_{p+1}]}')_{j-p_i+1}- g(\bm{y}_{[p_i,p_{p+1}]}')_{j-p_i+1},
    \end{align*}
    where the last equality holds since $x_{p_i}'=y_{p_i}'=\beta_i$.
    By Equation (\ref{eq:g'}), we get $g(\bm{x}')_{p_{i}} - g(\bm{y}')_{p_{i}} \equiv 0 \pmod{q}$.
    Moreover, by Lemmas \ref{lem:del'} and \ref{lem:burst-del}, we can conclude that either all entries in $\bm{g}(\bm{x}_{[p_i,p_{i+1}]}') - \bm{g}(\bm{y}_{[p_i,p_{i+1}]}')$ lie within $[0,q]$ or all lie within $[-q,0]$. 
    Then, we can obtain $\sigma \big(\bm{g}(\bm{x}')_{[p_i+1,p_{i+1}]} - \bm{g}(\bm{y}')_{[p_i+1,p_{i+1}]} \big)=1$.
    It follows by Lemma \ref{lem:number} that 
    \begin{align*}
      \sigma\big( \bm{g}(\bm{x}')- \bm{g}(\bm{y}') \big)
      \leq \sum_{i=0}^{m-1} \sigma\big( \bm{g}(\bm{x}')_{[p_i+1,p_{i+1}]} - \bm{g}(\bm{y}')_{[p_i+1,p_{i+1}]} \big)=m.
    \end{align*}
    
    Now we show that $\big| \mathrm{VT}^k\big(\bm{g}(\bm{x})-\bm{g}(\bm{y})\big) \big|\leq mq\sum_{i=1}^n i^k-m$ for $k\geq 0$.
    For $i\in [0,m]$, let $\bm{z}^{(i)}= \bm{y}^{(1)}\cdots \bm{y}^{(i)}\bm{x}^{(i+1)}\cdots \bm{x}^{(m)}$.  
    It follows that $\bm{z}^{(0)}=\bm{x}$, $\bm{z}^{(m)}=\bm{y}$, and $\mathcal{D}_1(\bm{z}^{(i)})\cap \mathcal{D}_1(\bm{z}^{(i+1)})\neq \emptyset$ when $q\geq 2$ and $\mathcal{D}_2(\bm{z}^{(i)})\cap \mathcal{D}_2(\bm{z}^{(i+1)})\neq \emptyset$ when $q=2$.
    We can compute
    \begin{align*}  
     \left| \mathrm{VT}^k\big(\bm{g}(\bm{x})-\bm{g}(\bm{y})\big) \right|
     &=\left| \mathrm{VT}^k\big(\bm{g}(\bm{x})\big) - \mathrm{VT}^k\big(\bm{g}(\bm{y})\big) \right|\\
     &=\left| \sum_{i=0}^{m-1}\mathrm{VT}^k\big(\bm{g}(\bm{z}^{(i)})\big) - \mathrm{VT}^k\big(\bm{g}(\bm{z}^{(i+1)})\big) \right|\\
     &\leq \sum_{i=0}^{m-1}\left| \mathrm{VT}^k \big(\bm{g}(\bm{z}^{(i)})\big) - \mathrm{VT}^k\big(\bm{g}(\bm{z}^{(i+1)})\big) \right|\\ 
     &\leq mq\sum_{i=1}^n i^k-m,
    \end{align*} 
    where the last inequality follows by Equations (\ref{eq:VT_del'}) and (\ref{eq:burst'}).
    This completes the proof.
\end{IEEEproof}

\begin{theorem}\label{thm:q}
  For $k\in [0,2s]$, let $n_k\triangleq q(2s+1)\sum_{i=1}^{n} i^k-2s-1$ and $a_k\in [0,n_k]$, we define the code
  \begin{align*}
    \mathcal{C}_{2}= \big\{\bm{x}\in \Sigma_q^n: \mathrm{VT}^k\big(\bm{g}(\bm{x})\big) \equiv a_{k} \pmod{n_k+1} \text{ for } k\in [0,2s]\big\}.
  \end{align*}
  By the pigeonhole principle, there exists a choice of parameters such that its redundancy is at most $(s+1)(2s+1)\log n+O(1)$ bits.
  Moreover, the following holds.
  \begin{itemize}
    \item If $q\geq 2$ and $s\geq 0$, $\mathcal{C}_{2}$ is a $q$-ary single-deletion $s$-substitution correcting code.
    \item If $q= 2$ and $s\geq 1$, $\mathcal{C}_{2}$ is a binary two-deletion $(s-1)$-substitution correcting code.
  \end{itemize}
\end{theorem}

\begin{IEEEproof}
    By considering all possible choices for $a_0,a_1,\ldots,a_{2s}$, we observe that the total number of codes $\mathcal{C}_2$ is $\prod_{k=0}^{2s}(m_k+1)$. 
    These codes partition the space $\Sigma_q^{n}$. 
    By the pigeonhole principle, there exists at least one choice of $a_0,a_1,\ldots,a_{2s}$ such that the code $\mathcal{C}_2$ contains at least $\frac{q^{n}}{\prod_{k=0}^{2s}(m_k+1)}= O\big(\frac{q^n}{n^{(s+1)(2s+1)}}\big)$ codewords. 
    The redundancy of such a code is at most $(s+1)(2s+1)\log n + O(1)$ bits.
    
    Now, we show that $\mathcal{C}_2$ is a $q$-ary single-deletion $s$-substitution correcting code for $q\geq 2$ and $s\geq 0$.
    Assume there are two sequences $\bm{x},\bm{y} \in \mathcal{C}_2$ such that $\mathcal{B}_{1,s}(\bm{x}) \cap \mathcal{B}_{1,s}(\bm{y})\neq \emptyset$.
    By Theorem \ref{thm:DS}, there exists some $m\in [1,2s+1]$ such that $\bm{x}=\bm{x}^{(1)}\bm{x}^{(2)}\cdots \bm{x}^{(m)}$ and $\bm{y}=\bm{y}^{(1)}\bm{y}^{(2)}\cdots \bm{y}^{(m)}$, where $|\bm{x}^{(i)}|=|\bm{y}^{(i)}|$ and $\mathcal{D}_{1}(\bm{x}^{(i)})\cap \mathcal{D}_{1}(\bm{y}^{(i)})\neq \emptyset$ for $i\in [1,m]$.
    Then by Lemma \ref{lem:ADS} and the condition that $\mathrm{VT}^k\big(\bm{g}(\bm{x})\big)\equiv \mathrm{VT}^k\big(\bm{g}(\bm{y})\big) \pmod{n_k+1}$, we have $\sigma\big(\bm{g}(\bm{x})-\bm{g}(\bm{y})\big)\leq m$ and $\mathrm{VT}^k\big(\bm{g}(\bm{x})-\bm{g}(\bm{y})\big)=0$ for $k\in [0,m-1]$.
    Finally, by Theorem \ref{thm:code}, $\mathcal{C}_{2}$ is a single-deletion $s$-substitution correcting code.
    
    It remains to show that $\mathcal{C}_2$ is a binary two-deletion $(s-1)$-substitution correcting code for $s \geq 1$.
    The proof follows with a similar discussion above and we include it here for completeness. 
    Assume there are two sequences $\bm{x},\bm{y} \in \mathcal{C}_2$ such that $\mathcal{B}_{2,s-1}(\bm{x}) \cap \mathcal{B}_{2,s-1}(\bm{y})\neq \emptyset$.
    By Theorem \ref{thm:DS}, there exists some $m\in [1,2s+1]$ such that $\bm{x}=\bm{x}^{(1)}\bm{x}^{(2)}\cdots \bm{x}^{(m)}$ and $\bm{y}=\bm{y}^{(1)}\bm{y}^{(2)}\cdots \bm{y}^{(m)}$, where $|\bm{x}^{(i)}|=|\bm{y}^{(i)}|$ and $\mathcal{D}_{2}(\bm{x}^{(i)})\cap \mathcal{D}_{2}(\bm{y}^{(i)})\neq \emptyset$ for $i\in [1,m]$.
    Then by Lemma \ref{lem:ADS} and the condition that $\mathrm{VT}^k\big(\bm{g}(\bm{x})\big)\equiv \mathrm{VT}^k\big(\bm{g}(\bm{y})\big) \pmod{n_k+1}$, we have $\sigma\big(\bm{g}(\bm{x})-\bm{g}(\bm{y})\big)\leq m$ and $\mathrm{VT}^k\big(\bm{g}(\bm{x})-\bm{g}(\bm{y})\big)=0$ for $k\in [0,m-1]$.
    Finally, by Theorem \ref{thm:code}, $\mathcal{C}_2$ is indeed a two-deletion $(s-1)$-substitution correcting code.
\end{IEEEproof}

\begin{remark}
    When $q\geq 3$ and $s=1$, the code presented in Theorem \ref{thm:q} can correct one deletion and one substitution, with a redundancy of at most $6\log n+O(1)$ bits.
    This method outperforms approaches based on syndrome compression or Linial's algorithm with pre-coding \cite{Sima-20-ISIT-tD,Song-22-IT-DS,Li-23-DS}, which requires a redundancy of $7\log n+o(\log n)$ bits. 
\end{remark}

\begin{remark}
  Since, by Remark \ref{rmk:g}, applying VT syndromes to accumulative differential sequences is essentially equivalent to applying them directly to differential sequences, when $ q \geq 3 $ and $ s=0 $, the code presented in Theorem \ref{thm:q} aligns with the construction of Nguyen et al. \cite{Nguyen-24-IT-1D}, requiring at most $ \log n + \log q $ bits of redundancy.
  Furthermore, by Remark \ref{rmk:run}, when $ q=2 $ and $ s=1 $, this code also coincides with the construction of Pi and Zhang \cite{Pi-24-arXiv-2E}, requiring at most $ 6\log n + O(1) $ bits of redundancy.
  Compared to \cite{Pi-24-arXiv-2E}, our proof is notably more concise.  
  Finally, it can be verified by Remark \ref{rmk:run} and Lemmas \ref{lem:del'} and \ref{lem:burst-del} that when $ q=2 $ and $ s=0 $, this code becomes a $^{\leq} 2$-burst-deletion correcting code, corresponding to the construction of Levenshtein \cite{Levenshtein-70-BD}, with a redundancy of at most $ \log n + 1 $ bits.
\end{remark}

\subsection{Encoder and Decoder}
Now we describe the encoding and decoding algorithms, but focus solely on the code presented in Theorem \ref{thm:t=1}, as the algorithms for the code constructed in Theorem \ref{thm:q} can be designed similarly.

\begin{definition}
For any $\bm{x}\in \Sigma_2^n$ and $k\in [0,2s]$, let $F^k(\bm{x})\triangleq \mathrm{VT}^{k}\big(\bm{f}(\bm{x})\big) \mod{\big((2s+1)\sum_{i=1}^n i^k+1 \big)}$ and $\bm{F}_b^k(\bm{x})$ be the binary representation of $F^k(\bm{x})$.
We define $\bm{F}_b(\bm{x})= \big(\bm{F}_b^0(\bm{x}), \bm{F}_b^1(\bm{x}),\ldots, \bm{F}_b^{2s}(\bm{x})\big)$.
\end{definition}

\begin{theorem}
  Let $\bm{F}_b(\cdot)$ be defined above and $\bm{R}_{2s+2}(\cdot)$ be the encoding function of the $(2s+2)$-fold repetition code, we define
  \begin{align*}
    \mathcal{E}(\boldsymbol{x})= \Big(\bm{x}, \bm{F}_b(\bm{x}), \bm{R}_{2s+2}\big( \bm{F}_b(\bm{F}_b(\bm{x}))\big) \Big).
  \end{align*}
  The code $\mathcal{C}= \{\mathcal{E}(\boldsymbol{x}): \boldsymbol{x}\in \Sigma_2^{n}\}$ is a single-deletion $s$-substitution correcting code with $(s+1)(2s+1)\log n+O(\log\log n)$ bits of redundancy. 
  Moreover, the encoding complexity is $O(n)$ and the brute force decoding complexity is $O(n^{s+2})$.
\end{theorem}

\begin{IEEEproof}
    For any $\boldsymbol{x}\in \Sigma_2^{n}$, let $n_1\triangleq |\bm{F}_b(\bm{x})|= (s+1)(2s+1)\log n+O(1)$ and $n_2\triangleq \big|\bm{R}_{2s+2}\big( \bm{F}_b(\bm{F}_b(\bm{x}))\big)\big|= O(\log\log n)$.
    Thus, the redundancy of $\mathcal{C}$ is $(s+1)(2s+1)\log n+O(\log\log n)$ bits.
    Note that $\bm{F}_b(\bm{x})$ and $\bm{R}_{2s+2}\big( \bm{F}_b(\bm{F}_b(\bm{x}))\big)$ can be computed in $O(n)$ and $O(n_1)=O(\log n)$ time, respectively, so the total encoding complexity is $O(n)$.
    Given any $\boldsymbol{y}\in \mathcal{B}_{1,s}\big(\mathcal{E}(\boldsymbol{x})\big)$, we have $\boldsymbol{y}_{[1,n-1]}\in \mathcal{B}_{1,s}(\bm{x})$, $\boldsymbol{y}_{[n+1,n+n_1-1]}\in \mathcal{B}_{1,s}\big(\bm{F}_b(\bm{x}) \big)$, and $\boldsymbol{y}_{[n+n_1+1,n+n_1+n_2-1]}\in \mathcal{B}_{1,s}\big(\bm{R}_{2s+2}( \bm{F}_b(\bm{F}_b(\bm{x}))) \big)$.
    Firstly, it can be easily verified that $\bm{R}_{2s+2}\big( \bm{F}_b(\bm{F}_b(\bm{x}))\big)$ can be recovered from $\boldsymbol{y}_{[n+n_1+1,n+n_1+n_2-1]}$ in $O(n_2)=O(\log\log n)$ time.
    Next, once $\bm{R}_{2s+2}\big( \bm{F}_b(\bm{F}_b(\bm{x}))\big)$ is recovered, we can compute $\bm{F}_b\big(\bm{F}_b(\bm{x})\big)$ in $O(n_2)=O(\log\log n)$ time. Then, applying Theorem \ref{thm:t=1}, $\bm{F}_b(\bm{x})$ can be recovered from $\boldsymbol{y}_{[n+1,n+n_1-1]}$ and $\bm{F}_b\big(\bm{F}_b(\bm{x})\big)$ in $O(n_1^{s+2})=O\big((\log n)^{s+2}\big)$ time via brute force.
    Finally, we can recover $\bm{x}$ from $\boldsymbol{y}_{[1,n-1]}$ and $\bm{F}_b(\bm{x})$ using brute force in $O(n^{s+2})$ time.
    Thus, $\mathcal{C}$ is a single-deletion $s$-substitution correcting code and the total decoding complexity is $O(n^{s+2})$.
    This completes the proof.
\end{IEEEproof}

\section{Discussions On Multiple-Deletion Correcting Codes}\label{sec:multiple}

In this section, we introduce potential tools to aid in correcting multiple deletions and multiple substitutions.

\subsection{Binary Codes}\label{subsec:binary}

As explained in Remark \ref{rmk:AS}, applying VT syndromes to accumulative sequences encounters limitations when correcting multiple deletions in binary alphabets. 
The limited effectiveness stems from the fact that the sign of $\sum_{i=1}^t x_i - \sum_{i=1}^t y_i$ cannot be determined solely by $\sum_{i=1}^t x_i$ when $t \geq 2$, where $x_i, y_i \in \Sigma_2$ for $i \in [1, t]$.
Observe that when $\sum_{i=1}^t y_i\leq 1$, the sign of $\sum_{i=1}^t x_i - \sum_{i=1}^t y_i$ becomes completely determined by $\sum_{i=1}^t x_i$. 
This motivates us to impose a constraint ensuring that the distance between consecutive ones in a sequence is at least $t$. 

\begin{definition}
   Let $t\geq 1$ be an integer, we say that $\bm{x}\in \Sigma_2^n$ is \emph{$t$-good} if the distance between two adjacent ones in it is at least $t$.
\end{definition}

\begin{lemma}\label{lem:multi-del}
    For a positive integer $t$, let $\bm{x}, \bm{y} \in \Sigma_2^n$ be distinct sequences such that $\mathcal{D}_t(\bm{x})\cap \mathcal{D}_t(\bm{y})\neq \emptyset$.
    If $\bm{x}$ and $\bm{y}$ are $t$-good, we have either $f(\bm{x})_{i}-f(\bm{y})_{i}\in \{0,1\}$ for $i\in [1,n]$ or $f(\bm{x})_{i}-f(\bm{y})_{i}\in \{-1,0\}$ for $i\in [1,n]$.
\end{lemma}

\begin{IEEEproof}
  Let $p_1$ and $p_2$ be the smallest and largest indices at which $\bm{x}$ and $\bm{y}$ differ, respectively.
  We first consider the case where $p_2-p_1\leq t-1$.
  By the definition of the accumulative sequence, we can compute
  \begin{equation*}
      f(\bm{x})_{i}-f(\bm{y})_{i}=
      \begin{cases}
          0, &\mbox{if } i\in [0,p_1-1];\\
          \sum_{j=p_1}^i x_j-\sum_{j=p_1}^i y_j, &\mbox{if } i \in [p_1, p_2];\\
          \sum_{j=p_1}^{p_2} x_j- \sum_{j=p_1}^{p_2} y_j, &\mbox{if } i \in [p_2+1,n].
      \end{cases}
  \end{equation*}
  Without loss of generality assume $x_{p_1}=1$ and $y_{p_1}=0$. 
  Since $\bm{y}$ is $t$-good, we get $\sum_{j=p_1}^{p_2}y_j \leq 1$.
  It then follows that $f(\bm{x})_{i}-f(\bm{y})_{i}\in \{1,0\}$ for $i\in [1,n]$.
  Thus, the conclusion holds for $p_2-p_1\leq t-1$.
  
  We now consider the case where $p_2-p_1\geq t$.
  Since $\mathcal{D}_t(\bm{x})\cap \mathcal{D}_t(\bm{y})\neq \emptyset$, there exists some $t'\in [1,t]$ such that either $\bm{x}_{[1,n] \setminus [p_1,p_1+t'-1]} = \bm{y}_{[1,n] \setminus [p_2-t'+1,p_2]}$ or $\bm{y}_{[1,n] \setminus [p_1,p_1+t'-1]} = \bm{x}_{[1,n] \setminus [p_2-t'+1,p_2]}$.
  Without loss of generality assume $\bm{x}_{[1,n] \setminus [p_1,p_1+t'-1]} = \bm{y}_{[1,n] \setminus [p_2-t'+1,p_2]}$.
  Let $p_x=p_1$ and $p_y=p_2-t'+1$, it follows that $\bm{x}_{[1,n] \setminus [p_x,p_x+t'-1]} = \bm{y}_{[1,n] \setminus [p_y,p_y+t'-1]}$ and $p_x<p_y$.
  By Figure \ref{fig:burst-deletion} and the definition of the accumulative sequence, we can compute 
  \begin{equation}\label{eq:multi-del}
      f(\bm{x})_{i}-f(\bm{y})_{i}=
      \begin{cases}
          0, &\mbox{if } i\in [0,p_x-1];\\
          \sum_{j=p_x}^i x_j-\sum_{j=p_1}^i y_j, &\mbox{if } i \in [p_x, p_x+t'-1];\\
          \sum_{j=p_x}^{p_x+t'-1} x_j-\sum_{j=i-t'+1}^i y_j, &\mbox{if } i \in [p_x+t', p_y+t'-1];\\
          \sum_{j=p_x}^{p_x+t'-1} x_j-\sum_{j=p_y}^{p_y+t'-1} y_j, &\mbox{if } i \in [p_y+t',n].
      \end{cases}
  \end{equation}
  We distinguish between two cases.
  \begin{itemize}
    \item If $x_{p_x}=1$ and $y_{p_x}=0$, since $\bm{x}$ and $\bm{y}$ are $t$-good, we get $\sum_{j=p_x}^{p_x+t'-1} x_j=x_{p_x}= 1$ and $\sum_{j=i-t'+1}^{i}y_j \leq 1$ for $i\geq p_x$.
        Then we can conclude that $f(\bm{x})_{i}-f(\bm{y})_{i}\in \{0,1\}$ for $i\in [1,n]$.
    \item If $x_{p_x}=0$ and $y_{p_x}=1$, since $x_{p_x+t'}=y_{p_x}=1$ and $\bm{x}$ and $\bm{y}$ are $t$-good, we get $\sum_{j=p_x}^{p_x+t'-1} x_j=0$ and $\sum_{j=i-t'+1}^{i}y_j \leq 1$ for $i\geq p_x$. Then we can conclude that $f(\bm{x})_{i}-f(\bm{y})_{i}\in \{-1,0\}$ for $i\in [1,n]$.
  \end{itemize}
  Thus, the conclusion also holds for $p_2-p_1\geq t$. This completes the proof.
\end{IEEEproof}

By repeating the discussions in Lemma \ref{lem:AS} and Theorem \ref{thm:t=1}, we can derive the following conclusion.

\begin{theorem}\label{thm:binary}
  For $k\in [0,2t+2s-2]$, where $t\geq 1$ and $s\geq 0$, let $n_k\triangleq (2t+2s-1)\sum_{i=1}^n i^k$ and $a_k\in [0,n_k]$, the code
  \begin{align*}
    \mathcal{C}_{3}= \big\{\bm{x} \text{ is $t$-good}: \mathrm{VT}^k\big(\bm{f}(\bm{x})\big) \equiv a_{k} \pmod{n_k+1} \text{ for } k\in [0,2t+2s-2]\big\}
  \end{align*}
  is a binary $t$-deletion $s$-substitution correcting code.
\end{theorem}
 
\begin{remark}
  In fact, sequences in which two adjacent ones are at least $ t $ positions apart have already been utilized to construct binary $ t $-deletion correcting codes in the literature \cite{Sima-20-IT-2D, Sima-21-IT-kD, Sima-20-ISIT-tD}.
  Here, we revisit this approach and clarify its underlying intuition within our framework.
\end{remark}

\subsection{Non-Binary Codes}\label{subsec:non-binary}

We now extend the above idea to non-binary alphabets. 
A natural generalization is to consider sequences in which the distance between any two consecutive non-zero entries is at least $ t $.
However, directly using such sequences is also not feasible, as the sign of $\sum_{j=p_x}^{p_x+t-1} x_j-\sum_{j=i-t+1}^i y_j$ (see in Equation (\ref{eq:multi-del})) may still remain undetermined by $ \sum_{j=p_x}^{p_x+t-1} x_j $.
Luckily, for reasons similar to those in Equations (\ref{eq:property}) and (\ref{eq:property'}), analyzing the differential sequences allows this problem to be resolved.

\begin{definition}
  Let $t\geq 1$ be an integer, we say that $\bm{x}\in \Sigma_q^n$ is \emph{$t$-valid} if the distance between two adjacent non-zero entries in it is at least $t+1$.
\end{definition}

\begin{observation}
  Let $\bm{x}\in \Sigma_q^n$ be a $t$-valid sequence, any substring of its differential sequence of length $t$ contains at most two non-zero entries.
  If such substring contains two non-zero entries, these two entries are consecutive.
  Moreover, for $i\in [1,n-t'+1]$ and $t'\in [1,t]$, it holds that 
  \begin{equation}\label{eq:valid}
    \sum_{j=i}^{i+t'-1} d(\bm{x})_j \leq q.
  \end{equation}
\end{observation}

\begin{remark}
  If we permit the distance between two adjacent non-zero entries in a sequence to be exactly $t$, Equation (\ref{eq:valid}) no longer holds.
\end{remark}

\begin{lemma}\label{lem:multi-del'}
    For a positive integer $t$, let $\bm{x}, \bm{y} \in \Sigma_q^n$ be distinct sequences such that $\mathcal{D}_t(\bm{x})\cap \mathcal{D}_t(\bm{y})\neq \emptyset$.
    If $\bm{x}$ and $\bm{y}$ are $t$-valid, we have either $g(\bm{x})_{i}-g(\bm{y})_{i}\in [0,q]$ for $i\in [1,n]$ or $g(\bm{x})_{i}-g(\bm{y})_{i}\in [-q,0]$ for $i\in [1,n]$.
    Moreover, it is clear that $|g(\bm{x})_{1}-g(\bm{y})_{1}|\leq q-1$.
\end{lemma}

\begin{IEEEproof}
  Let $p_1$ and $p_2$ be the smallest and largest indices at which $\bm{x}$ and $\bm{y}$ differ, respectively, we get $x_i=y_i$ for $i\not\in [p_1,p_2]$.
  We first consider the case where $p_2-p_1\leq t-1$, then $\bm{x}_{[1,n] \setminus [p_1,p_2]} = \bm{y}_{[1,n] \setminus [p_1,p_2]}$.
  By the definition of the accumulative differential sequence, we can compute
  \begin{equation*}
      g(\bm{x})_{i}-g(\bm{y})_{i}=
      \begin{cases}
          0, &\mbox{if } i\in [0,p_1-1];\\
          \sum_{j=p_1}^i d(\bm{x})_j-\sum_{j=p_1}^i d(\bm{y})_j, &\mbox{if } i \in [p_1, p_2];\\
          \sum_{j=p_1}^{p_2+1} d(\bm{x})_j- \sum_{j=p_1}^{p_2+1} d(\bm{y})_j, &\mbox{if } i \in [p_2+1,n].
      \end{cases}
  \end{equation*}
  Without loss of generality assume $x_{p_1}> y_{p_1}$, since $x_{p_1-1}=y_{p_1-1}$, $x_{p_2+1}=y_{p_2+1}$, and $\bm{x}$ and $\bm{y}$ are $t$-valid, we have $x_{p_1-1}=y_{p_1-1}=0$, $x_{p_1+1}=\cdots=x_{p_2+1}=y_{p_2+1}=0$, and there is at most one non-zero entry in $\bm{y}_{[p_1,p_2]}$.
  By the definition of the differential sequence, we get $d(\bm{x})_{p_1}=x_{p_1}>y_{p_1}=d(\bm{y})_{p_1}$, $\sum_{i=p_1}^{p_2+1}= \sum_{i=p_1}^{p_1+1} d(\bm{x})_i=q$, and $\sum_{i=p_1}^{p_2+1}= d(\bm{y})_i\leq q$.
  This implies that $g(\bm{x})_{i}-g(\bm{y})_{i}\in [0,q]$ for $i\in [1,n]$.
  Thus, the conclusion holds for $p_2-p_1\leq t-1$.
   
  We now consider the case where $p_2-p_1\geq t$. 
  Since $\mathcal{D}_t(\bm{x})\cap \mathcal{D}_t(\bm{y})\neq \emptyset$, there exists some $t'\in [1,t]$ such that either $\bm{x}_{[1,n] \setminus [p_1,p_1+t'-1]} = \bm{y}_{[1,n] \setminus [p_2-t'+1,p_2]}$ or $\bm{y}_{[1,n] \setminus [p_1,p_1+t'-1]} = \bm{x}_{[1,n] \setminus [p_2-t'+1,p_2]}$.
  Without loss of generality assume $\bm{x}_{[1,n] \setminus [p_1,p_1+t'-1]} = \bm{y}_{[1,n] \setminus [p_2-t'+1,p_2]}$.
  Let $p_x=p_1$ and $p_y=p_2-t'+1$, it follows that $\bm{x}_{[1,n] \setminus [p_x,p_x+t'-1]} = \bm{y}_{[1,n] \setminus [p_y,p_y+t'-1]}$ and $p_x<p_y$.
  By Figure \ref{fig:burst-deletion} and the definition of the differential sequence, we get
  $d(\bm{x})_{p_y+t'}\equiv \sum_{j=p_y}^{p_y+t'} d(\bm{y})_j\pmod{q}$.
  Since $\bm{y}$ is $t$-valid, by Equation (\ref{eq:valid}), we have
  \begin{equation}\label{eq:valid'}
    d(\bm{x})_{p_y+t'}- \sum_{j=p_y}^{p_y+t'} d(\bm{y})_j\in \{-q,0\}.
  \end{equation}
  By Figure \ref{fig:burst-deletion'} and the definition of the accumulative differential sequence, we can compute 
  \begin{equation*}
      g(\bm{x})_{i}-g(\bm{y})_{i}=
      \begin{cases}
          0, &\mbox{if } i\in [0,p_x-1];\\
          \sum_{j=p_x}^i d(\bm{x})_j-\sum_{j=p_x}^i d(\bm{y})_j, &\mbox{if } i \in [p_x, p_x+t'-1];\\
          \sum_{j=p_x}^{p_x+t'} d(\bm{x})_j-d(\bm{y})_{p_x}-\sum_{j=i-t'+1}^i d(\bm{y})_j, &\mbox{if } i \in [p_x+t', p_y+t'-1];\\
          \sum_{j=p_x}^{p_x+t'} d(\bm{x})_j-d(\bm{y})_{p_x}+d(\bm{x})_{p_y+t'}-\sum_{j=p_y}^{p_y+t'} d(\bm{y})_j, &\mbox{if } i \in [p_y+t',n].
      \end{cases}
  \end{equation*}
  We further distinguish between two cases.
    \begin{itemize}
        \item If $x_{p_x}>y_{p_x}$, since $x_{p_x-1}=y_{p_x-1}$, $x_{p_x+t'}=y_{p_x}$, and $\bm{x}$ is $t$-valid, we have $x_{p_x-1}=y_{p_x-1}=0$ and $x_{p_x+1}=\cdots= x_{p_x+t'}=y_{p_x}=0$.
            By the definition of the differential sequence, we get $d(\bm{x})_{p_x}=x_{p_x}>0$, $d(\bm{y})_{p_x}=y_{p_x}=0$, and $\sum_{j=p_x}^{p_x+t'} d(\bm{x})_j= \sum_{i=p_x}^{p_x+1} d(\bm{x})_{j}=q$.
            Further by Equations (\ref{eq:valid}) and (\ref{eq:valid'}), we can conclude that $g(\bm{x})_{i}-g(\bm{y})_{i}\in [0,q]$ for $i\in [1,n]$.

        \item If $x_{p_x}<y_{p_x}$, since $x_{p_x-1}=y_{p_x-1}$, $x_{p_x+t'}=y_{p_x}$, and $\bm{x}$ is $t$-valid, we have $x_{p_x-1}=y_{p_x-1}=0$ and $x_{p_x}=\cdots x_{p_x+t'-1}=0$. By the definition of the differential sequence, we can compute $\sum_{i=p_x}^{p_x+t'} d(\bm{x})_i=d(\bm{y})_{p_x}$.
            Further by Equations (\ref{eq:valid}) and (\ref{eq:valid'}), we get $g(\bm{x})_{i}-g(\bm{y})_{i}\in [-q,0]$ for $i\in [1,n]$.
    \end{itemize}
    Thus, the conclusion also holds for $p_2-p_1\geq t$. This completes the proof.
\end{IEEEproof}  

By repeating the discussions in Lemma \ref{lem:ADS} and Theorem \ref{thm:q}, we can derive the following conclusion.

\begin{theorem}\label{thm:non-binary}
  For $k\in [0,2t+2s-2]$, let $n_k\triangleq q(2t+2s-1)\sum_{i=1}^n i^k$ and $a_k\in [0,n_k]$, the code
  \begin{align*}
    \mathcal{C}_4= \big\{\bm{x} \text{ is $t$-valid}: \mathrm{VT}^k\big(\bm{g}(\bm{x})\big) \equiv a_{k} \pmod{n_k+1} \text{ for } k\in [0,2t+2s-2]\big\}
  \end{align*}
  is a $q$-ary $t$-deletion $s$-substitution correcting code.
\end{theorem}

\begin{remark}
  While the approach used in Theorems \ref{thm:binary} and \ref{thm:non-binary} is valuable for constructing error-correcting codes, the scarcity of good and valid sequences means that the redundancies of the codes presented in these theorems are sufficiently large. 
  Therefore, simply applying this strategy is insufficient.
  It is necessary to introduce additional functions that map sequences into good or valid sequences. For good sequences, Sima et al. \cite{Sima-20-IT-2D, Sima-20-ISIT-tD, Sima-21-IT-kD} have proposed effective methods to achieve this. Exploring strategies for transforming sequences into valid sequences remains a promising direction for future research.
\end{remark}

\section{conclusion}\label{sec:conclusion}

In this paper, we introduce a novel technique to analyze a pair of sequences whose error balls intersect non-trivially, establishing a clear connection between error correction and burst-deletion correction. Consequently, it is sufficient to focus on correcting a burst of deletions. By leveraging existing strategies in the field of burst-deletion correction, we construct codes capable of correcting both deletions and substitutions, with some constructions matching existing results and others surpassing current methods.

\begin{appendix}[Extensions to Adjacent Transpositions]

We now extend the partitioning technique to handle errors involving $ t $ deletions combined with $ s $ substitutions and adjacent transpositions (transpositions for short), where a transposition swaps two neighboring symbols.
For any $\bm{x}\in \Sigma_2^n$, let $\mathcal{B}_{t,s}^{D,ST}(\bm{x})$ denote the set of all sequences obtainable from $\bm{x}$ after $t$ deletions combined with $ s $ substitutions and transpositions.

\begin{theorem}\label{thm:DST}
  For integers $ t\geq 1 $ and $ s\geq 0 $, let $ \bm{x}, \bm{y} \in \Sigma_2^n $ be distinct sequences such that $ \mathcal{B}_{t,s}^{D,ST}(\bm{x}) \cap \mathcal{B}_{t,s}^{D,ST}(\bm{y}) \neq \emptyset $. Then, there exists a partition
  \begin{gather*}
  \bm{x} = \bm{x}^{(1)} \bm{x}^{(2)} \cdots \bm{x}^{(m)},\\
  \bm{y} = \bm{y}^{(1)} \bm{y}^{(2)} \cdots \bm{y}^{(m)},
  \end{gather*}
  with $ m \leq 2t + 2s - 1 $, such that for each $ i \in [1, m] $, $ |\bm{x}^{(i)}| = |\bm{y}^{(i)}| $ and $ \mathcal{D}_{t}(\bm{x}^{(i)}) \cap \mathcal{D}_{t}(\bm{y}^{(i)}) \neq \emptyset$.
\end{theorem}

Before proceeding to the proof, we present the following conclusions, which will be utilized later.

\begin{observation}\label{obs:par}
  Let $ \bm{x}, \bm{z} \in \Sigma_2^n $ be such that $ \mathcal{D}_{t}(\bm{x}) \cap \mathcal{D}_{t}(\bm{z}) \neq \emptyset $, where $ t \ge 1 $. For $i\in [1,n-1]$, it holds that $ \mathcal{D}_{t}(\bm{x}_{[1,i]}) \cap \mathcal{D}_{t}(\bm{z}_{[1,i]}) \neq \emptyset$ and $ \mathcal{D}_{t}(\bm{x}_{[i+1,n]}) \cap \mathcal{D}_{t}(\bm{z}_{[i+1,n]}) \neq \emptyset$.
\end{observation}

\begin{lemma}\label{lem:DST}
  Let $ \bm{x}=\bm{x}^{(1)} \bm{x}^{(2)}, \bm{z}=\bm{z}^{(1)} \bm{z}^{(2)} \in \Sigma_2^n $ be such that $|\bm{x}^{(i)}|=|\bm{z}^{(i)}|$ and $ \mathcal{D}_{t}(\bm{x}^{(i)}) \cap \mathcal{D}_{t}(\bm{z}^{(i)}) \neq \emptyset $ for $i\in \{1,2\}$, where $ t \ge 1 $. If $ \bm{y} $ can be obtained from $ \bm{z} $ after a transposition at position $ p_y\leq n'\triangleq |\bm{x}^{(1)}| $, then $ \bm{x} $ and $ \bm{y} $ can be partitioned into $ \bm{x} = \tilde{\bm{x}}^{(1)} \tilde{\bm{x}}^{(2)} \tilde{\bm{x}}^{(3)} $ and $ \bm{y} = \tilde{\bm{y}}^{(1)}\tilde{\bm{y}}^{(2)}\tilde{\bm{y}}^{(3)} $, such that for each $i\in \{1,2,3\}$, $ |\tilde{\bm{x}}^{(i)}| = |\tilde{\bm{y}}^{(i)}|$ and $ \mathcal{D}_{t}(\tilde{\bm{x}}^{(i)}) \cap \mathcal{D}_{t}(\tilde{\bm{y}}^{(i)}) \neq \emptyset$.
\end{lemma}

\begin{IEEEproof}
    If $\bm{y}=\bm{z}$, the conclusion is straightforward.
    Thus, we assume $\bm{y}\neq \bm{z}$, then $y_{p_y}\neq y_{p_y+1}$.
    Since $\mathcal{D}_{t}(\bm{x}^{(1)})\cap \mathcal{D}_{t}(\bm{z}^{(1)})\neq \emptyset$, there exists some $t'\in [1,t]$ and $p_x,p_z\in [1,n'-t'+1]$ such that $\bm{x}_{[1,n']\setminus [p_x,p_x+t'-1]}^{(1)}= \bm{z}_{[1,n']\setminus [p_z,p_z+t'-1]}^{(1)}$.
    By symmetry, without loss of generality assume $p_x\leq p_z$.
    We distinguish between three cases.

    \begin{itemize}
        \item If $p_y\leq p_z-1$, let 
            \begin{gather*}
                \tilde{\bm{x}}^{(1)}= \bm{x}_{[1,p_y+1]},~\tilde{\bm{x}}^{(2)}=\bm{x}_{[p_y+2,n']},\tilde{\bm{x}}^{(3)}=\bm{x}_{[n'+1,n]},\\
                \tilde{\bm{y}}^{(1)}= \bm{y}_{[1,p_y+1]},~\tilde{\bm{y}}^{(2)}=\bm{y}_{[p_y+2,n']},\tilde{\bm{y}}^{(3)}=\bm{y}_{[n'+1,n]}.
            \end{gather*}
            By Observation \ref{obs:par}, we have $ \mathcal{D}_{t}(\tilde{\bm{x}}^{(i)}) \cap \mathcal{D}_{t}(\tilde{\bm{y}}^{(i)}) \neq \emptyset$ for $t\in \{2,3\}$.
            We further distinguish between two cases.
        \begin{itemize}
            \item If $p_y\leq p_x-1$, recall that $y_{p_y}\neq y_{p_y+1}$, we have $ \mathcal{D}_{1}(\bm{x}_{[p_y,p_y+1]}) \cap \mathcal{D}_{1}(\bm{y}_{[p_y,p_y+1]}) \neq \emptyset$.
                Since $\bm{y}_{[1,p_y-1]}=\bm{z}_{[1,p_y-1]}=\bm{x}_{[1,p_y-1]}$, it follows that $ \mathcal{D}_{1}(\tilde{\bm{x}}^{(1)}) \cap \mathcal{D}_{1}(\tilde{\bm{y}}^{(1)}) \neq \emptyset$.

            \item If $p_x\leq p_y\leq p_z-1$, we consider the value of $t'$.
            \begin{itemize}
                \item If $t'\geq 2$, let $\ell= \min\{t',p_y-p_x+2\}$, we have $\bm{x}_{[1,p_y+1]\setminus [p_x,p_x+\ell-1]}= \bm{y}_{[1,p_y+1]\setminus [p_y-\ell+2,p_y+1]}$. 
                    This implies that $ \mathcal{D}_{t'}(\tilde{\bm{x}}^{(1)}) \cap \mathcal{D}_{t'}(\tilde{\bm{y}}^{(1)}) \neq \emptyset$.

                \item If $t'=1$, we have $\bm{x}_{[1,p_y+1]\setminus \{p_x\}}= \bm{y}_{[1,p_y+1]\setminus \{p_y\}}$. This implies that $ \mathcal{D}_{1}(\tilde{\bm{x}}^{(1)}) \cap \mathcal{D}_{1}(\tilde{\bm{y}}^{(1)}) \neq \emptyset$.
            \end{itemize}
        \end{itemize}
        Consequently, this partition is the desired one.

        \item If $p_z\leq p_y\leq n'-1$, let 
        \begin{gather*}
            \tilde{\bm{x}}^{(1)}=\bm{x}_{[1,p_y-1]},~\tilde{\bm{x}}^{(2)}=\bm{x}_{[p_y,n']}, ~\tilde{\bm{x}}^{(3)}=\bm{x}_{[n'+1,n]},\\
            \tilde{\bm{y}}^{(1)}=\bm{y}_{[1,p_y-1]},~\tilde{\bm{y}}^{(2)}=\bm{y}_{[p_y,n']}, ~\tilde{\bm{y}}^{(3)}=\bm{y}_{[n'+1,n]}.
        \end{gather*}
        By Observation \ref{obs:par}, we have $ \mathcal{D}_{t}(\tilde{\bm{x}}^{(i)}) \cap \mathcal{D}_{t}(\tilde{\bm{y}}^{(i)}) \neq \emptyset$ for $t\in \{1,3\}$.
        Recall that $y_{p_y}\neq y_{p_y+1}$, we have $ \mathcal{D}_{1}(\bm{x}_{[p_y,p_y+1]}) \cap \mathcal{D}_{1}(\bm{y}_{[p_y,p_y+1]}) \neq \emptyset$.
        Since $\bm{y}_{[p_y+2,n']}=\bm{z}_{[p_y+2,n']}=\bm{x}_{[p_y+2,n']}$, it follows that $ \mathcal{D}_{1}(\tilde{\bm{x}}^{(2)}) \cap \mathcal{D}_{1}(\tilde{\bm{y}}^{(2)}) \neq \emptyset$.
        Consequently, this partition is the desired one.

        \item If $p_y=n'$, let 
        \begin{gather*}
            \tilde{\bm{x}}^{(1)}=\bm{x}_{[1,n'-1]},~\tilde{\bm{x}}^{(2)}=\bm{x}_{[n',n'+1]},~\tilde{\bm{x}}^{(3)}=\bm{x}_{[n'+2,n]},\\
            \tilde{\bm{y}}^{(1)}=\bm{y}_{[1,n'-1]},~\tilde{\bm{y}}^{(2)}=\bm{y}_{[n',n'+1]},~\tilde{\bm{y}}^{(3)}=\bm{y}_{[n'+2,n]}.
        \end{gather*}
        By Observation \ref{obs:par}, we have $ \mathcal{D}_{t}(\tilde{\bm{x}}^{(i)}) \cap \mathcal{D}_{t}(\tilde{\bm{y}}^{(i)}) \neq \emptyset$ for $t\in \{1,3\}$.
        Recall that $y_{p_y}\neq y_{p_y+1}$, we have $\mathcal{D}_1(\tilde{\bm{x}}^{(2)})\cap \mathcal{D}_1(\tilde{\bm{y}}^{(2)})\neq \emptyset$.
        Consequently, this partition is the desired one.
    \end{itemize}
    In each of these cases, we can obtain a desired partition, which completes the proof.
\end{IEEEproof}

For binary alphabets, Gabrys et al. \cite[Lemma 4]{Gabrys-18-IT-DT} demonstrated that the order of deletions and transpositions is commutative with respect to the final sequence. 
By symmetry, the order of insertions and transpositions is also commutative. 
Consequently, the combined order of deletions, insertions, and substitution-transpositions does not affect the final outcome.

We are now ready to prove Theorem \ref{thm:DST}.

\begin{IEEEproof}[Proof of Theorem \ref{thm:DST}]
    Since the order of insertions, deletions, and substitution-transpositions is commutative with respect to the final outcome, the sequence $\bm{y}$ can be obtained from $ \bm{x} $ after $ t $ deletions, then $t$ insertions, followed by $ s'\leq 2s $ substitution-transpositions. 
    We prove the theorem by induction on $s'$.
    
    For the base case where $s'=0$, the conclusion follows by Lemma \ref{lem:D}.
    Now, assuming that the conclusion is valid for $s'<2s$, we examine the scenario where $s'=2s$.
    Observe that there exists some sequence $\bm{z}$ obtainable from $ \bm{x} $ through the same deletions and insertions, but with $ s'-1 $ substitution-transpositions, such that $ \bm{y} $ can be obtained from $ \bm{z} $ by a single substitution-transposition, by applying the induction hypothesis on the pair $(\bm{x},\bm{z})$ and by using Lemmas \ref{lem:DS} and \ref{lem:DST}, we can conclude that the conclusion is also valid for $s'=2s$.
    This completes the proof.
\end{IEEEproof}
\end{appendix}

\end{document}